\documentstyle[a4,12pt,epsf,cite]{article}
\newcommand{\be}{\begin{equation}}
\newcommand{\ee}{\end{equation}}
\newcommand{\bea}{\begin{eqnarray}}
\newcommand{\eea}{\end{eqnarray}}
\newcommand{\egg}{$\eta \rightarrow \gamma \gamma \;$}
\newcommand{\epg}{$\eta \rightarrow \pi^0 \gamma \gamma\;$}
\newcommand{\egm}{$\eta \rightarrow \gamma \mu^- \mu^+ \;$}
\newcommand{\ege}{$\eta \rightarrow \gamma e^- e^+ \;$}
\newcommand{\egl}{$\eta \rightarrow \gamma \, l^- l^+ \;$}
\newcommand{\sla}{\hspace{-0.5 em}/}
\newdimen\gap \gap=2pt
\newdimen\gaq \gaq=4pt
\newdimen\columnheight
\newdimen\columndepth
\newcommand{\col}[1]{\setbox255=\hbox{#1}
               \columnheight=\ht255 \columndepth=\dp255
               \advance\columnheight by \gaq
               \advance\columndepth by \gaq
               {\vrule height\columnheight depth\columndepth width0pt}#1}
\begin{document}
\baselineskip 1.5pc
\begin{center}
{\Large \bf Strong $U_A(1)$ breaking in radiative $\eta$ decays}
\vskip 10mm
M. Takizawa\footnote{E-mail address: takizawa@ins.u-tokyo.ac.jp}\\
{\it Institute for Nuclear Study, University of Tokyo,} \\
{\it Tanashi, Tokyo 188, Japan}\\
\vskip 5mm
Y. Nemoto\footnote{E-mail address: nemoto@th.phys.titech.ac.jp} and 
M. Oka\footnote{E-mail address: oka@th.phys.titech.ac.jp}\\
{\it Department of Physics, Tokyo Institute of Technology,} \\
{\it Meguro, Tokyo 152, Japan} 
\end{center}
\vskip 5mm
\begin{abstract}
\baselineskip 1.5pc
We study the \egg, \egm and \epg decays using an extended three-flavor 
Nambu-Jona-Lasinio model that includes the 't~Hooft instanton induced 
interaction.  
We find that the $\eta$-meson mass, the \egg, \egm and \epg  
decay widths are in good agreement with the experimental values when the 
$U_{A}(1)$ breaking is strong and the flavor $SU(3)$ singlet-octet 
mixing angle $\theta$ is about zero.  
The calculated $\eta \gamma \gamma^\ast$ transition form factor has 
somewhat weaker dependence on the squared four-momentum of the virtual photon. 
The effects of the $U_A(1)$ anomaly on the scalar quark contents in the 
nucleon, the $\Sigma_{\pi N}$ and $\Sigma_{KN}$ terms and the baryon 
number one and two systems are also studied.
\end{abstract}
\section{Introduction}
\hspace*{\parindent}It is well known that the QCD action has an approximate 
$U_L(3) \times U_R(3)$ chiral symmetry and its sub-symmetry, $U_A(1)$
symmetry, is explicitly broken by the anomaly.  The $U_A(1)$ symmetry
breaking is manifested in the heavy mass of the $\eta'$ meson. 
The physics of the $\eta$ and $\eta'$ mesons have been extensively
studied in the $1/N_C$ expansion approach \cite{tHooft74}. In the $N_C
\to \infty$ limit, the $U_A(1)$ anomaly is turned off and then the
$\eta$ meson becomes degenerate with the pion and the $\eta'$ meson
becomes a pure $\bar ss$ state with $m_{\eta'}^2 (N_C \to \infty) \simeq 
2 m_K^2 - m_\pi^2 \simeq (687 \,{\rm MeV})^2$ \cite{Vene79}.  So the
$U_A(1)$ anomaly pushes up $m_{\eta}$ by about 400MeV and $m_{\eta'}$
by about 300MeV. It means that not only the $\eta'$ meson but also the
$\eta$ meson is largely affected by the $U_A(1)$ anomaly. 
\par
    In order to understand the role of the $U_A(1)$ anomaly in the
low-energy QCD, it may be important to study the $\eta$-meson decays
as well as its mass and decay constant.  Among the $\eta$-meson
decays, \egg, \egl ($l$ denotes $e$ and/or $\mu$) and \epg decays are 
interesting.  
They have no final state interactions and involve only neutral mesons
so that the electromagnetic transitions are induced only by the
internal (quark) structure of the mesons.
\par
The \egg decay is related to the Adler-Bell-Jackiw (ABJ) triangle anomaly 
\cite{ABJ69} through the partial conservation of axialvector current 
(PCAC) hypothesis.
One of the useful and widely used frameworks for studying the phenomena 
related to the axial-vector anomaly and the spontaneous chiral symmetry 
breaking is the chiral effective meson lagrangian given by Wess and Zumino 
\cite{WZ71} and developed by Witten \cite{Wit83}.  
The $\eta,\eta' \to \gamma \gamma$ decays have been studied using the 
Wess-Zumino-Witten (WZW) lagrangian with the corrections at one-loop order 
in the chiral perturbation and it has been shown that the two-photon decay 
widths can be explained with the $\eta$-$\eta'$ mixing angle 
$\theta \simeq -20^\circ$ \cite{DHL85}.
From the chiral perturbation theory (ChPT) \cite{GL84} point of view, 
the WZW term is derived in the chiral limit and is of order $p^4$.  
As discussed in \cite{GL85}, to reliably calculate $SU(3)$ breaking effects 
of the $\eta, \eta' \to 2 \gamma$ decays, the low-energy expansion to order 
$p^6$ has to be carried out.  However in \cite{DHL85} full analysis of order 
$p^6$ has not been performed.  Furthermore, because of the $U_A(1)$ anomaly, 
the singlet channel decay amplitude $\eta_0 \to \gamma \gamma$ derived using 
PCAC + ABJ anomaly should be modified so as to become 
renormalisation group invariant \cite{SV92}.
\par
The \egl decay is closely related to the \egg decay since it is considered as
$\eta \to \gamma \gamma^\ast \to \gamma l^+ l^-$.  By observing the 
muon pair invariant mass square spectrum of the \egl decay, one is able to 
obtain the transition form factor for the $\eta \gamma \gamma^\ast$ vertex.
It gives us the information of the size of the $\eta$-meson.  In \cite{Ame92-1}
the transition form factors for the $\pi^0 \gamma \gamma^\ast$, 
$\eta \gamma \gamma^\ast$ and $\eta' \gamma \gamma^\ast$ vertices have been 
studied in the vector meson dominance model, the constituent quark loop model, 
the QCD-inspired interpolation model by Brodsky-Lepage and the ChPT. 
However none of the models have taken into account the effects of the 
$U_A(1)$ anomaly explicitly. 
\par
Unlike the \egg decay, there is no low-energy theorem for the \epg decay 
and therefore it is not trivial.
In ChPT \cite{Ame,Jett,Ko}, there is no lowest order 
$O(p^{2})$ contribution to the \epg process because the involved mesons 
are neutral. Likewise the next order $O(p^{4})$ tree diagrams  do not exist. 
Thus the $O(p^{4})$ one-loop diagrams give the leading term in this process, 
but the contribution is two orders of magnitude smaller than the 
experimental value. This is because the pion loop violates the G-parity 
invariance and the kaon loop is also suppressed by the large kaon mass.
The $O(p^{6})$ contributions are dominant and the result is a factor two 
smaller than the experimental value.
Although these results based on ChPT are not too far from the experimental 
value, it is noted that the higher order $O(p^{6})$
terms in the perturbation expansion are larger than the leading $O(p^{4})$
terms and the results contain ambiguous parameters that cannot be determined
well from other processes.
\par
    The purpose of this paper is to study the \egg, \egl and \epg decays in 
the framework of a chiral quark model so that the quark structure of the 
$\eta$ meson is explicitly taken into account.  In such a model the explicit 
chiral symmetry breaking by the current quark masses can be included in a 
nonperturbative way.  The effects of the $U_A(1)$ anomaly can also be 
represented by the coupling of light quarks to the instanton as was pointed 
out by 't Hooft \cite{tHooft76}.  We here take the three-flavor 
Nambu-Jona-Lasinio (NJL) model as the chiral quark model.  
The model involves the $U_L(3) \times U_R(3)$ symmetric 
four-quark interaction and the six-quark flavor-determinant interaction 
\cite{tHooft76,KKM71} incorporating effects of the $U_A(1)$ anomaly.  It is
widely used in recent years to study such topics as the quark condensates 
in vacuum, the spectrum of low-lying mesons, the flavor-mixing properties 
of the low-energy hadrons, etc. \cite{KH88,HK94,BJM88,KKT88,RA88,KLVW90}.  
In this approach the explicit chiral symmetry breaking 
and the $U_A(1)$ anomaly on the \egg, \egl and \epg decay amplitudes 
can be calculated consistently with those on the $\eta$-meson mass, $\eta$ 
decay constant and mixing angle within the model applicability.
Furthermore one is able to study how the $\eta$-meson properties 
change when the strength of the $U_A(1)$ breaking interaction is changed in 
this approach.  
\par
We have studied the \egg decay \cite{TO95} and 
the \epg decay \cite{NOT96} in this approach and found that these decay widths
are reproduced when the $U_A(1)$ breaking interaction is much stronger than 
the previous studies in the three-flavor 
NJL model \cite{KH88,HK94,BJM88,KKT88,RA88,KLVW90}.
The $U_A(1)$ breaking six-quark flavor-determinant interaction induced by 
the instanton \cite{tHooft76} gives rise to flavor mixing not only in the 
pseudoscalar channel but also in the scalar channel. It is further argued that
the instanton can play an important role in description of spin-spin forces, 
particularly for light baryons \cite{SR89,OT89,MT93}.  
Since the $U_A(1)$ breaking interaction is found to be rather strong, 
it is important to reexamine the effects of the  $U_A(1)$ breaking interaction
on the scalar quark contents in the nucleon, the $\Sigma_{\pi N}$ and 
$\Sigma_{KN}$ terms and the baryon number one and two systems. 
\par
The paper is organized as follows.  In sect. 2 we explain methods for 
calculating the $\eta$-meson mass, mixing angle and decay constant in the 
three-flavor NJL model.  We describe the calculations of the \egg, \egm 
and \epg decay amplitudes in sect. 3.  The numerical results 
of the $\eta$-meson decays are presented in sect. 4.  We study the $\bar uu$, 
$\bar dd$ and $\bar ss$ contents in the nucleon and the $\Sigma_{\pi N}$ 
and $\Sigma_{KN}$ terms in sect. 5.  Sect. 6 is devoted to the study of the 
effects of the $U_A(1)$ breaking interaction on the baryon number one and 
two systems. Finally, summary and concluding remarks are given in sect. 7.
\section{$\eta$-meson in the three-flavor NJL model}
\hspace*{\parindent}We work with the following NJL model lagrangian density:
\bea
{\cal L} & = & {\cal L}_0 + {\cal L}_4 + {\cal L}_6 , \label{njl1} \\
{\cal L}_0 & = & \bar \psi \,\left( i \partial_\mu \gamma^\mu - \hat m 
\right) \, \psi \, ,
\label{njl2} \\
{\cal L}_4 & = & {G_S \over 2} \sum_{a=0}^8 \, \left[\, \left( 
\bar \psi \lambda^a \psi \right)^2 + \left( \bar \psi \lambda^a i \gamma_5 
\psi \right)^2 \, \right] \, ,
\label{njl3} \\
{\cal L}_6 & = & G_D \left\{ \, {\rm det} \left[ \bar \psi_i (1 - \gamma_5) 
\psi_j \right] + {\rm det}  \left[ \bar \psi_i (1 + \gamma_5) \psi_j 
\right] \, \right\} \, .
\label{njl4}
\eea
Here the quark field $\psi$ is a column vector in color, flavor and Dirac 
spaces and $\lambda^a (a=0\ldots 8)$ is the $U(3)$ generator in flavor space. 
The free Dirac lagrangian ${\cal L}_0$ incorporates the current quark mass 
matrix $\hat m = {\rm diag}(m_u, m_d, m_s)$ which breaks the chiral 
$U_L(3) \times U_R(3)$ invariance explicitly. ${\cal L}_4$ is a QCD 
motivated four-fermion interaction, which is chiral $U_L(3) \times U_R(3)$
invariant.  The 't Hooft determinant ${\cal L}_6$ represents the $U_A(1)$
anomaly.  It is a $3 \times 3$ determinant with respect to flavor with 
$i,j = {\rm u,d,s}$.    
\par
    Quark condensates and constituent quark masses are self-consistently 
determined by the gap equations
\bea
M_u & = & m_u - 2 G_S \langle \bar u u \rangle - 2 G_D \langle \bar d d 
\rangle \langle \bar s s \rangle \, , \nonumber \\
M_d & = & m_d - 2 G_S \langle \bar d d \rangle - 2 G_D \langle \bar s s 
\rangle \langle \bar u u \rangle \, , \nonumber \\
M_s & = & m_s - 2 G_S \langle \bar s s \rangle - 2 G_D \langle \bar u u 
\rangle \langle \bar d d \rangle \, , \label{gap}
\eea
with
\be
\langle \bar a a \rangle = - {\rm Tr}^{(c,D)} \, \left[ i S_F^a (x = 0) \right]
=  - \int^{\Lambda} \frac{d^4p}{(2 \pi)^4} {\rm Tr}^{(c,D)} \left[ 
\frac{i}{p_{\mu} \gamma^{\mu} - M_a + i \varepsilon} \right] \, . \label{cond}
\ee
Here the covariant cutoff $\Lambda$ is introduced to regularize the 
divergent integral and ${\rm Tr}^{(c,D)}$ 
means trace in color and Dirac spaces.
\par
    The pseudoscalar channel quark-antiquark scattering amplitude
\be
\langle p_3 , \bar p_4 ; {\rm out} | p_1 , \bar p_2 ; {\rm in} \rangle = 
(2 \pi)^4 \delta^4(p_3 + p_4 - p_1 - p_2) {\cal T}_{q \bar q}
\ee
is then calculated in the ladder approximation. We assume the isospin 
symmetry too.  In the $\eta$ and $\eta'$ channel, the explicit expression is 
\be
{\cal T}_{q \bar q} = -
\left(
\begin{array}{c} 
\bar u(p_3) \lambda^8 i \gamma_5 v(p_4) \\
\bar u(p_3) \lambda^0 i \gamma_5 v(p_4) 
\end{array} 
\right)^T \,
\left(
\begin{array}{cc}
A(q^2) & B(q^2) \\
B(q^2) & C(q^2) \\
\end{array}
\right) \,
\left(
\begin{array}{c}
\bar v(p_2) \lambda^8 i \gamma_5 u(p_1) \\
\bar v(p_2) \lambda^0 i \gamma_5 u(p_1)
\end{array}
\right) \, , \label{qas1}
\ee
with 
\bea
A(q^2) & = & \frac{2}{{\rm det}{\bf D}(q^2)} 
\left\{ 2 ( G_0 G_8 - G_m G_m ) I^0 (q^2) - G_8 \right\} \, , \label{qas2} \\
B(q^2) & = & \frac{2}{{\rm det}{\bf D}(q^2)}
\left\{- 2 ( G_0 G_8 - G_m G_m ) I^m (q^2) - G_m \right\} \, , \label{qas3} \\
C(q^2) & = & \frac{2}{{\rm det}{\bf D}(q^2)}
\left\{ 2 ( G_0 G_8 - G_m G_m ) I^8 (q^2) - G_0 \right\} \, , \label{qas4} 
\eea
and 
\bea
G_0 & = & \frac{1}{2} G_S - \frac{1}{3} ( 2 \langle \bar uu \rangle + 
\langle \bar ss \rangle ) G_D \, , \\
G_8 & = & \frac{1}{2} G_S - \frac{1}{6} ( \langle \bar ss \rangle - 4  
\langle \bar uu \rangle ) G_D \, , \\
G_m & = & - \frac{1}{3 \sqrt{2}} ( \langle \bar ss \rangle - 
\langle \bar uu \rangle ) G_D \, . 
\eea
The quark-antiquark bubble integrals are
\bea
I^0(q^2) & = & i \int^{\Lambda} \frac{d^4p}{(2 \pi)^4} {\rm Tr}^{(c,f,D)}
\left[ S_F(p) \lambda^0 i \gamma_5 S_F(p+q) \lambda^0 i \gamma_5 \right]
\, , \label{int1} \\
I^8(q^2) & = & i \int^{\Lambda} \frac{d^4p}{(2 \pi)^4} {\rm Tr}^{(c,f,D)}
\left[ S_F(p) \lambda^8 i \gamma_5 S_F(p+q) \lambda^8 i \gamma_5 \right]
\, , \label{int2} \\
I^m(q^2) & = & i \int^{\Lambda} \frac{d^4p}{(2 \pi)^4} {\rm Tr}^{(c,f,D)}
\left[ S_F(p) \lambda^0 i \gamma_5 S_F(p+q) \lambda^8 i \gamma_5 \right]
\, , \label{int3} 
\eea
with $q = p_1 + p_2$.  The $2 \times 2$ matrix ${\bf D}$ is 
\be
{\bf D}(q^2) = 
\left( 
\begin{array}{cc}
D_{11}(q^2) & D_{12}(q^2) \\
D_{21}(q^2) & D_{22}(q^2) 
\end{array}
\right) \, , \label{mat}
\ee
with
\bea
D_{11}(q^2) & = & 2 G_8 I^8(q^2) + 2 G_m I^m(q^2) - 1 \, , \label{mat11}\\
D_{12}(q^2) & = & 2 G_8 I^m(q^2) + 2 G_m I^0(q^2) \, , \label{mat12} \\
D_{21}(q^2) & = & 2 G_0 I^m(q^2) + 2 G_m I^8(q^2) \, , \label{mat21} \\
D_{22}(q^2) & = & 2 G_0 I^0(q^2) + 2 G_m I^m(q^2) - 1 \, . \label{mat22}
\eea
{}From the pole position of the scattering amplitude Eq. (\ref{qas1}), the 
$\eta$-meson mass $m_{\eta}$ is determined.
\par
    The scattering amplitude Eq. (\ref{qas1}) can be diagonalized by rotation
in the flavor space 
\bea
{\cal T}_{q \bar q} & = & -
\left(
\begin{array}{c} 
\bar u(p_3) \lambda^8 i \gamma_5 v(p_4) \\
\bar u(p_3) \lambda^0 i \gamma_5 v(p_4) 
\end{array} 
\right)^T   {\bf T}_{\theta}^{-1} {\bf T}_{\theta} 
\left(
\begin{array}{cc}
A(q^2) & B(q^2) \\
B(q^2) & C(q^2) \\
\end{array}
\right) {\bf T}^{-1}_{\theta}  \nonumber \\
&& \times {\bf T}_{\theta}  
\left(
\begin{array}{c}
\bar v(p_2) \lambda^8 i \gamma_5 u(p_1) \\
\bar v(p_2) \lambda^0 i \gamma_5 u(p_1)
\end{array}
\right) \, , \label{qasm1} \\
& = & -
\left(
\begin{array}{c} 
\bar u(p_3) \lambda^{\eta} i \gamma_5 v(p_4) \\
\bar u(p_3) \lambda^{\eta'} i \gamma_5 v(p_4) 
\end{array} 
\right)^T \, 
\left(
\begin{array}{cc}
D^{\eta}(q^2) & 0 \\
0 & D^{\eta'}(q^2) 
\end{array}
\right) \nonumber \\
&& \times
\left(
\begin{array}{c}
\bar v(p_2) \lambda^{\eta} i \gamma_5 u(p_1) \\
\bar v(p_2) \lambda^{\eta'} i \gamma_5 u(p_1)
\end{array}
\right) \, , \label{qasm2}
\eea
with $\lambda^{\eta} \equiv \cos \theta  \lambda^8 - \sin \theta \lambda^0$,
$\lambda^{\eta'} \equiv \sin \theta  \lambda^8 + \cos \theta \lambda^0$ and
\be
{\bf T}_{\theta} = \left(
\begin{array}{cc}
\cos \theta & -\sin \theta \\
\sin \theta & \cos \theta 
\end{array}
\right) \, .
\ee
The rotation angle $\theta$ is determined by 
\be
\tan 2 \theta = \frac{2 B(q^2)}{C(q^2) - A(q^2)} \, . \label{angle}
\ee
So $\theta$ depends on $q^2$.  At $q^2 = m_{\eta}^2$,  $\theta$ represents the 
mixing angle of the $\lambda^8$ and $\lambda^0$ components in the $\eta$-meson
state.  In the usual effective pseudoscalar meson lagrangian approaches, the 
$\eta$ and $\eta'$ mesons are analyzed using the $q^2$-independent 
$\eta$-$\eta'$ mixing angle.  Because of the $q^2$-dependence, $\theta$ cannot 
be interpreted as the $\eta$-$\eta'$ mixing angle.  The origin of the 
$q^2$-dependence is that the $\eta$ and $\eta'$ meson have the internal 
quark structures.
\par
For the $\eta'$ meson, since the NJL model 
does not confine quarks, the $\eta'$-meson state has the unphysical imaginary 
part which corresponds to the $\eta' \to q \bar q$ decays. 
Therefore we do not apply our model to the 
$\eta'$ meson in this article.
\par
    We define the effective $\eta$-quark 
coupling constant $g_{\eta}$ by introducing 
additional vertex lagrangian,
\be
  {\cal L}_{\eta q q} = g_{\eta} \overline{\psi} i\gamma_{5}\lambda^{\eta}
\psi \phi_{\eta} \, ,
\ee
with $ \lambda^{\eta}=\cos \theta \lambda^{8}-\sin \theta \lambda^{0}$.
Here $\phi$ is an auxiliary meson field introduced for convenience and 
the effective $\eta$-quark coupling constant is calculated from the residue 
of the $q \bar{q}$-scattering amplitudes at the $\eta$ pole, i.e. 
$g_{\eta}^2 = \lim_{q^2 \to m_{\eta}^2} (q^2 - m_{\eta}^2) D_{\eta}(q^2)$
The $\eta$ decay constant $f_{\eta}$ is determined by calculating the 
quark-antiquark one-loop graph, 
\be
f_{\eta} = \frac{g_{\eta}}{m_{\eta}^2} \int^{\Lambda} \frac{d^4p}{(2 \pi)^4}
{\rm Tr}^{(c,f,D)} \bigl[q^{\mu} \gamma_{\mu} \gamma_5 
\frac{\lambda^{\eta}}{2}
S_F(p) i \gamma_5 \lambda^{\eta} S_F(p-q)\bigr] \left\vert_{q^2 = m_{\eta}^2} 
\right. \, . \label{edc} 
\ee
\par
    One can easily show that in the $U_A(1)$ limit, i.e., $G_D = 0$ and 
$m_{u,d} \not= m_s$, the $\eta$ meson becomes the ideal mixing state composed 
of u and d-quarks, namely, $m_{\eta} = m_{\pi}$, $g_{\eta} = g_{\pi}$, 
$f_{\eta} = f_{\pi}$ and $\tan \theta = - \sqrt{2}$.
\section{$\eta$-meson decays}
\subsection{\egg decay}
\hspace*{\parindent}The  $P \to \gamma \gamma$ ($P = \pi^0, \eta, \eta'$) 
decay amplitude is given by 
\be
\langle \gamma (k_1) \gamma (k_2) \vert P (p) \rangle = i (2 \pi)^4 
\delta^4 (k_1 + k_2 - p) \varepsilon_{\mu \nu \rho \sigma} 
\epsilon^{\mu}_1 \epsilon^{\nu}_2 k_1^{\rho} k_2^{\sigma} 
\widetilde{\cal T}_{P \to \gamma \gamma}(p^2) \, , \label{ampl}
\ee
where $\epsilon_1$ and $\epsilon_2$ are the polarization vectors of the 
photon and the $P \to \gamma \gamma$ decay width 
$\Gamma(P \to \gamma \gamma)$ is given by
\be
\Gamma(P \to \gamma \gamma) = 
\left\vert \widetilde{\cal T}_{P \to \gamma \gamma} \right\vert^2 
\frac{m_P^3}{64 \pi} \, .
\ee
\par
    In the analysis of the $\eta$-$\eta'$ mixing angle, the 
$\eta, \eta' \to \gamma \gamma$ decay widths and the current algebra formulae 
are used \cite{GK87}.  In order to see the assumptions used in the derivation 
of the current algebra formulae for the $\eta, \eta' \to \gamma \gamma$ 
decay amplitudes, we re-derive them here. 
\par
    The starting points are the following PCAC relations which are modified 
by the ABJ anomaly. 
\bea
\partial^\mu A_\mu^8 \, & = &
\, f_{8 \eta} m_\eta^2 \phi_\eta + f_{8 \eta'} m_{\eta'}^2 \phi_{\eta'} 
 - {\alpha \over 4\pi} N_c D^8 \varepsilon_{\mu \nu \rho \sigma} 
   F^{\mu \nu} F^{\rho \sigma} \quad , \label{PCAC8} \\
\partial^\mu A_\mu^0 \, & = &
\, f_{0 \eta} m_\eta^2 \phi_\eta + f_{0 \eta'} m_{\eta'}^2 \phi_{\eta'} 
 - {\alpha \over 4\pi} N_c D^0 \varepsilon_{\mu \nu \rho \sigma} 
   F^{\mu \nu} F^{\rho \sigma} \quad . \label{PCAC0}
\eea 
with
\be
\begin{array}{cc}
\langle 0 \vert \partial^\mu A_\mu^8 \vert \eta \rangle = 
f_{8 \eta} m_\eta^2 \, , \quad & \quad
\langle 0 \vert \partial^\mu A_\mu^8 \vert \eta' \rangle = 
f_{8 \eta'} m_{\eta'}^2 \, , \\
\langle 0 \vert \partial^\mu A_\mu^0 \vert \eta \rangle = 
f_{0 \eta} m_\eta^2 \, , \quad & \quad
\langle 0 \vert \partial^\mu A_\mu^0 \vert \eta' \rangle = 
f_{0 \eta'} m_{\eta'}^2 \, , \\
\end{array}
\label{dcmix}
\ee
and $D^a$ ($a=3,8,0$) is defined as ${\rm tr} [ \{Q,Q\} \frac{\lambda^a}{2}]$ 
with $Q \equiv \frac{1}{2}( \lambda^3 + \frac{1}{\sqrt{3}} \lambda^8 )$.
$\alpha$ is the fine structure constant of QED and 
$F^{\mu \nu}$ is the electromagnetic field tensor.  The simple 
$\eta$-$\eta'$ state mixing is then assumed, i.e.,  
\be
\left\{
  \begin{array}{l}
\vert \eta \rangle \, = 
\cos \theta \, \vert \eta_8 \rangle 
- \sin \theta \, \vert \eta_0 \rangle \, , \\
\vert \eta' \rangle =
\sin \theta \, \vert \eta_8 \rangle 
+ \cos \theta \, \vert \eta_0 \rangle \, , 
  \end{array} \right.
\label{etamix}
\ee
and
\be
\langle 0 \vert \partial^\mu A_\mu^8 \vert \eta_8 \rangle = 
f_{8} m_{\eta_8}^2 \, , \quad  \quad
\langle 0 \vert \partial^\mu A_\mu^0 \vert \eta_0 \rangle = 
f_{0} m_{\eta_0}^2 \, . \label{sodc}
\ee
From Eqs. (\ref{dcmix})-(\ref{sodc}), $f_{8\eta} = f_8 \cos \theta$, 
$f_{8\eta'} = f_8 \sin \theta$, $f_{0\eta} = - f_0 \sin \theta$ and 
$f_{0\eta'} = f_0 \cos \theta$ are obtained.  
The $\eta$-meson field $\phi_\eta$ is expressed as follows.
\bea
\phi_\eta & = & \frac{1}{m_\eta^2} \left( \frac{\cos \theta}{f_8} 
\partial^\mu A_\mu^8 - \frac{\sin \theta}{f_0} \partial^\mu A_\mu^0 \right) 
\nonumber \\
&& + \frac{\alpha}{4\pi} \varepsilon_{\mu \nu \rho \sigma} 
F^{\mu \nu} F^{\rho \sigma} \frac{1}{m_\eta^2}
\left( \frac{\cos \theta}{f_8} \frac{1}{2\sqrt{3}}
- \frac{\sin \theta}{f_0} \sqrt{\frac{2}{3}} \right) \, . \label{PCACeta}
\eea
Using the LSZ-reduction formula, the \egg decay amplitude is given by
\bea
T_{\eta \to \gamma \gamma} & \equiv & \langle \gamma (k_1) \gamma (k_2) \vert
\eta (p) \rangle \nonumber \\
& = & i \int d^4 x e^{-i p \cdot x} ( \Box + m_\eta^2 ) 
\langle \gamma (k_1) \gamma (k_2) \vert \phi_\eta (x) \vert 0 \rangle \, .
\eea
By inserting modified PCAC relation Eq. (\ref{PCACeta}),
\be
T_{\eta \to \gamma \gamma} = \lim_{p^2 \to m_\eta^2} \left[\,
T_{\eta \to \gamma \gamma}^{(1)}(p^2) + 
T_{\eta \to \gamma \gamma}^{(2)}(p^2) \, \right] \, ,
\ee
with 
\bea
T_{\eta \to \gamma \gamma}^{(1)}(p^2) & = & \frac{(m_\eta^2 - p^2)}{m_\eta^2} 
i \int d^4 x e^{-i p \cdot x} \nonumber \\
&& \times \Big \langle \gamma (k_1) \gamma (k_2) \Big \vert
{\rm T} \left[ \frac{\cos \theta}{f_8} \partial^\mu A_\mu^8 - 
\frac{\sin \theta}{f_0} \partial^\mu A_\mu^0 \right] \Big \vert 0 \Big \rangle 
\, , \\
T_{\eta \to \gamma \gamma}^{(2)}(p^2) & = & \frac{(m_\eta^2 - p^2)}{m_\eta^2} 
i \int d^4 x e^{-i p \cdot x}\, 
\left( \frac{\cos \theta}{f_8} \frac{1}{2\sqrt{3}}
- \frac{\sin \theta}{f_0} \sqrt{\frac{2}{3}} \right) \nonumber \\
&& \times \Big \langle \gamma (k_1) \gamma (k_2) \Big \vert
{\rm T} \left[ \frac{\alpha}{4\pi} \varepsilon_{\mu \nu \rho \sigma} 
F^{\mu \nu} F^{\rho \sigma}  \right] 
\Big \vert 0 \Big \rangle \, .
\eea
Using the current algebra techniques and from the consideration of the general
structure of the matrix element, we can show that 
$T_{\eta \to \gamma \gamma}^{(1)} (p^2 = 0) = 0$ \cite{IZ80}.
$T_{\eta \to \gamma \gamma}^{(2)}$ can be calculated easily,
\bea
T_{\eta \to \gamma \gamma}^{(2)}(p^2) & = & 
i \, (2\pi)^4 \,\delta^4 (k_1 + k_2 - p)\, 
\varepsilon_{\mu \nu \rho \sigma} \epsilon^{\mu}_1 \epsilon^{\nu}_2 
k_1^{\rho} k_2^{\sigma} \,\frac{(m_\eta^2 - p^2)}{m_\eta^2} \nonumber \\
&& \quad \quad \times \frac{\alpha}{\pi} \, \frac{1}{\sqrt{3}}
\left( \frac{\cos \theta}{f_8} - 2 \sqrt{2} \frac{\sin \theta}{f_0} 
\right) \, ,
\eea
therefore the reduced invariant amplitude 
$\widetilde{\cal T}_{\eta \to \gamma \gamma}$ in the soft $\eta$ limit is 
\be
\widetilde{\cal T}_{\eta \to \gamma \gamma} (p^2 = 0) = \frac{\alpha}{\pi} \, 
\frac{1}{\sqrt{3}} \left( \frac{\cos \theta}{f_8} 
- 2 \sqrt{2} \frac{\sin \theta}{f_0} \right) \, . \label{etaggpcac}
\ee
Similarly, the reduced invariant amplitude 
$\widetilde{\cal T}_{\eta' \to \gamma \gamma}$ in the soft $\eta'$ limit is 
\be
\widetilde{\cal T}_{\eta' \to \gamma \gamma} (p^2 = 0) = \frac{\alpha}{\pi} \,
\frac{1}{\sqrt{3}} \left( \frac{\sin \theta}{f_8} 
+ 2 \sqrt{2} \frac{\cos \theta}{f_0} \right) \, . \label{etpggpcac}
\ee
Finally, $\widetilde{\cal T}_{\eta \to \gamma \gamma} (p^2 = 0) \simeq 
\widetilde{\cal T}_{\eta \to \gamma \gamma} (p^2 = m_\eta^2)$
and $\widetilde{\cal T}_{\eta' \to \gamma \gamma} (p^2 = 0) \simeq 
\widetilde{\cal T}_{\eta' \to \gamma \gamma} (p^2 = m_{\eta'}^2)$ 
are assumed in the usual analysis of the $\eta$-$\eta'$ mixing angle 
\cite{GK87}.
Since $m_\eta$ and $m_{\eta'}$ are not significantly smaller than the 
typical hadronic mass scale, the above results should not be taken 
quantitatively. 
Furthermore, because of the ABJ anomaly by the gluons ($U_A(1)$ anomaly) 
in the flavor singlet channel, the saturation of the flavor singlet 
axialvector current by the $\eta$ and $\eta'$ field is rather questionable.
\begin{figure}[t]
\begin{center}
\leavevmode
\epsfxsize=300pt
\epsfbox{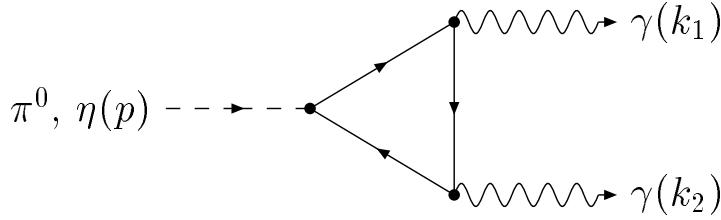}    
\caption{The quark triangle diagram for $\pi^0, \, \eta \to \gamma \gamma$ 
decays.}
\label{fig:e2ggf}
\end{center}
\end{figure}
\par
    In the NJL model, the on meson mass shell $\pi^0, \eta \to \gamma \gamma$
decay amplitudes can be obtained by calculating the quark triangle diagrams 
shown in Fig. \ref{fig:e2ggf} and our results are
\bea
\widetilde{\cal T}_{\pi^0 \to \gamma \gamma}(p^2 = m_{\pi^0}^2) 
& = & \frac{\alpha}{\pi} g_{\pi} F_{\pi^0}^u \, , \label{amplpi}  \\ 
\widetilde{\cal T}_{\eta \to \gamma \gamma}(p^2 = m_\eta^2)
& = & \frac{\alpha}{\pi} g_{\eta}
\frac{1}{3\sqrt{3}}  \Big[ \cos \theta \left\{ 5 F_\eta^u - 2 F_\eta^s 
\right\}\nonumber\\
& & \qquad \qquad 
- \sin \theta \sqrt{2} \left\{ 5 F_\eta^u + F_\eta^s \right\}
\Big] \, .  \label{ampleta}
\eea
Here $F_P^a$ ($a=u,s$ and $P=\pi^0,\eta$) is defined as
\bea
F_P^a & = &\int_0^1 dx \int_0^1 dy \,
         \frac{2 (1-x) M_a}{M_a^2 - m_P^2 x (1-x) (1-y)} \nonumber \\
& =  & \frac{4 M_a}{m_P^2} \, \arcsin^2 \left( \frac{m_P}{2M_a} \right) \, . 
\label{fam}
\eea
We can see that the integrand of $F_P^a$ has an unphysical pole when 
$m_P \geq 2 M_a$.  It is due to lack of the confinement mechanism in the NJL 
model.
\par
    In the chiral limit, the pion mass vanishes and $F_{\pi^0}^u$ becomes 
$1/M_u$.  In this limit, the Goldberger-Treiman (GT) relation at the quark 
level, $M_u = g_{\pi} f_{\pi}$, holds in the NJL model and this leads to 
$\widetilde{\cal T}_{\pi^0 \to \gamma \gamma} = \alpha/(\pi f_{\pi})$ which 
is same as the tree-level results in the Wess-Zumino-Witten lagrangian 
approach \cite{WZ71,Wit83}.  It should be 
mentioned that we have to integrate out the triangle diagrams without 
introducing a cutoff $\Lambda$ in order to get the above result  though the 
cutoff is introduced in the gap equations in the NJL model.  
In the $U(3)_L \times U(3)_R$ version of the NJL model, 
the WZW term has been derived using the bosonization method with 
the heat-kernel expansion \cite{ER86,W89}.  In their approach, $O(1/\Lambda)$
term has been neglected and it is equivalent to taking the $\Lambda
\to \infty$ limit. 
\subsection{\egl decays}
\hspace*{\parindent}The $P \to \gamma \, l^- l^+$ ($P = \pi^0, \eta$ and 
$l = e, \mu$) decay amplitude is given by 
\bea
\langle \gamma (k) \, l^-(q_1) l^+(q_2) \vert P(p) \rangle & = & 
i (2 \pi)^4 \delta^4 (q_1 + q_2 + k -p) \nonumber \\
&& \times \varepsilon_{\mu \nu \rho \sigma} \,
e \bar u(q_1) \gamma^\mu v(q_2) \frac{1}{q^2} k^\rho q^\sigma \,
 \widetilde{\cal T}_{P \to \gamma \, l^- l^+}(q^2) \, , 
\eea
where $q \equiv q_1 + q_2$ and $u$ and $v$ denotes lepton and antilepton's 
spinors.  The lepton pair invariant mass square spectrum is 
\bea
\frac{d \, \Gamma(P \to \gamma \, l^- l^+)}{dx} & = & 
\frac{\alpha}{96 \pi^2}\,  m_P^3 \, \frac{(1-x)^3}{x} 
\left( 1 - \frac{r^2}{x} \right)^{\frac{1}{2}}
 \left( 1 + \frac{r^2}{2x} \right) \nonumber \\ 
&& \times 
\left\vert \widetilde{\cal T}_{P \to \gamma \, l^- l^+}(m_P^2 x) 
\right\vert^2 \, , \quad {\rm with} \quad (1 \geq x \geq r^2)\, . 
\label{p2gllsp} 
\eea
Here $x \equiv q^2/m_P^2$ and $r \equiv 2 m_l /m_P$ and the total 
$P \to \gamma \, l^- l^+$ decay width is given by 
\bea
&& \Gamma(P \to \gamma \, l^- l^+) \nonumber \\  
&& = \frac{\alpha}{96 \pi^2}\,  m_P^3 \,
\int_{r^2}^1  dx \frac{(1-x)^3}{x} 
\left( 1 - \frac{r^2}{x} \right)^{\frac{1}{2}}
\left( 1 + \frac{r^2}{2x} \right)  
\left\vert \widetilde{\cal T}_{P \to \gamma \, l^- l^+}(m_P^2 x) 
\right\vert^2 \, . \quad
\eea 
\begin{figure}[t]
\begin{center}
\leavevmode
\epsfxsize=300pt
\epsfbox{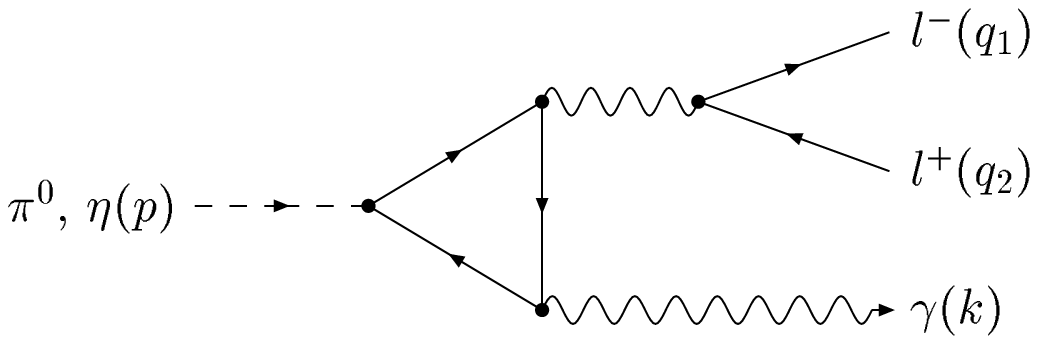}    
\caption{The quark triangle diagram for $\pi^0, \eta \to \gamma \, l^- l^+$ 
decays.}
\label{fig:e2gllf}
\end{center}
\end{figure}
\par
    By calculating the diagram shown in Fig. \ref{fig:e2gllf}, we obtain the 
$\pi^0, \eta \to \gamma \, l^- l^+$ decay amplitudes as follows. 
\bea
\widetilde{\cal T}_{\pi^0 \to \gamma \, l^- l^+}(q^2) & = & 
\frac{\alpha}{\pi} g_{\pi} G_\pi^u(q^2)  \label{p2gllamp}\\
\widetilde{\cal T}_{\eta \to \gamma \, l^- l^+}(q^2) & = & 
\frac{\alpha}{\pi} g_{\eta} \frac{1}{3\sqrt{3}}  
\Big[ \cos \theta \left\{ 5 G_\eta^u(q^2) - 2 G_\eta^s(q^2) 
\right\}\nonumber\\
& & \qquad \qquad 
- \sin \theta \sqrt{2} \left\{ 5 G_\eta^u(q^2) + G_\eta^s(q^2) \right\}
\Big] \, ,  \label{e2gllamp}
\eea
where 
\bea
G_P^a(q^2) & = &\int_0^1 dx \int_0^1 dy \,
         \frac{2 (1-x) M_a}{M_a^2 - m_P^2 x (1-x) (1-y) - q^2 x (1 -x) y} 
\nonumber \\
& =  & \frac{4 M_a}{(m_P^2 - q^2)} \, 
\left\{ \arcsin^2 \left( \frac{m_P}{2M_a} \right) - 
        \arcsin^2 \left( \frac{q}{2M_a} \right) \right\} \, , 
\label{inte2gll}
\eea
with $q \equiv \sqrt{q^2}$.  We can find the following relations, 
$G_P^a(q^2 = 0) = F_P^a$ and 
$\widetilde{\cal T}_{P \to \gamma \, l^- l^+}(q^2 = 0) = 
\widetilde{\cal T}_{P \to \gamma \gamma}$.
\par
    From the observed data for the two-photon transition 
$\gamma \gamma^* \to P$ and the lepton pair invariant mass spectrum of the 
$P \to \gamma l^- l^+$ decay, one can obtain the $P \gamma \gamma^*$ 
transition form factor $f_{P \gamma \gamma^*}(q^2)$ defined by 
\be
  f_{P \gamma \gamma^*}(q^2) \equiv 
\frac{\widetilde{\cal T}_{P \to \gamma \, l^- l^+}(q^2)}
{\widetilde{\cal T}_{P \to \gamma \, l^- l^+}(q^2 = 0)} \, .
\ee
For the spacelike $q^2$, $G^a_P(q^2)$ is given by
\be
G_P^a(q^2)  = \frac{2 M_a}{(m_P^2 - q^2)} \,  
\left\{2 \arcsin^2 \left( \frac{m_P}{2M_a} \right) + 
\frac{1}{2} \ln^2 \frac{\beta + 1}{\beta - 1} \right\} \, , 
\ee
with $\beta = \sqrt{1 - 4 M_a^2/q^2}$.
We introduce the slope 
parameter $\Lambda_P$ by 
\be
\frac{1}{\Lambda_P^2} \equiv \left .\frac{d}{dq^2} f_{P \gamma \gamma^*}(q^2)
\right \vert_{q^2 = 0} \equiv \frac{r_P^2}{6} \, , \label{lamdap}
\ee
and $\Lambda_P$ corresponds to the pole mass if one fits the 
$q^2$-dependence of $f_{P \gamma \gamma^*}(q^2)$ by means of a single-pole 
term.  In the case of the charge form factor $f_c(q^2)$ of the charged 
pseudoscalar meson, 
$6/ \Lambda_c^2 \equiv 6 d f_c(q^2) / dq^2 = \langle r_c^2 \rangle$ is the 
mean square charge radius of the charged pseudoscalar meson.  Therefore, 
it is natural to consider that $\Lambda_P$ is related to the size of the 
neutral pseudoscalar meson $P$. 
\par
    $\Lambda_\pi$ can be calculated easily in the chiral limit.  Using 
Eqs. (\ref{p2gllamp}), (\ref{inte2gll}) and (\ref{lamdap}), we get 
$\Lambda_\pi = \sqrt{12} M_u$.  On the other hand, 
the NJL model predicts $\Lambda_c = 2 \sqrt{2} f_\pi$ for the pion in the 
chiral limit \cite{BM88}.  Since $\Lambda_\pi$ is expressed in terms of the 
dynamical quantity of the model: $M_u$ in contrast with $\Lambda_c$ which is 
expressed in terms of the observed quantity: $f_\pi$, the $P \gamma \gamma^*$
transition form factor may be more sensitive to the dynamical structure of the
pseudoscalar meson than the charge form factor.
\subsection{\epg decay}
\begin{figure}[t]
\begin{center}
\leavevmode
\epsfxsize=300pt
\epsfbox{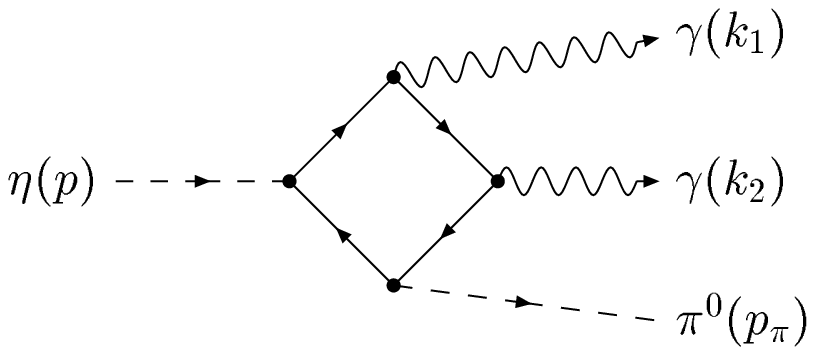}    
\caption{The quark box diagram for \epg decay.}
\label{fig:e2pggf}
\end{center}
\end{figure}

\hspace*{\parindent}The \epg decay amplitude is given by
\be
  \langle \pi^{0}(p_{\pi}) \gamma(k_{1},\epsilon_{1}) \gamma(k_{2},
\epsilon_{2}) | \eta(p) \rangle = i (2\pi)^{4} \delta^{4}
(p_{\pi}+k_{1}+k_{2}-p) \epsilon_{1}^{\mu} \epsilon_{2}^{\nu} T_{\mu\nu}\, .
\ee
The dominant contributions to this process in this model are the quark-box 
diagrams given in Fig. \ref{fig:e2pggf}.
Following the evaluation of the quark-box diagrams performed in \cite{Ng93},
we obtain
\be
  T_{\mu\nu}=-i \frac{1}{\sqrt{3}}(\cos \theta -\sqrt{2} \sin \theta )
e^{2}g_{\eta}g_{\pi} \int \frac{d^{4}q}{(2\pi)^{4}} 
\sum_{i=1}^{6} U_{\mu\nu}^{i}, \label{epgg1}
\ee
with
\begin{eqnarray}
  U_{\mu\nu}^{1} & = & {\rm tr} \left\{ \gamma_{5} \frac{1}{q \sla 
- M + i\epsilon} \gamma_{5} \frac{1}{q \sla + p \sla - k \sla_{1}  
- k \sla_{2} - M +i\epsilon} \right. \nonumber \\
  & \times & \left. \gamma_{\nu} \frac{1}{q \sla + p \sla - k \sla_{1} 
- M + i\epsilon} \gamma_{\mu} \frac{1}{q \sla + p \sla - M +i\epsilon} 
\right\} , \label{umn1}\\
  U_{\mu\nu}^{2} & = & {\rm tr} \left\{ \gamma_{5} \frac{1}{q \sla 
- M + i\epsilon} \gamma_{5} \frac{1}{q \sla + k \sla_{2} - M +i\epsilon} 
\right. \nonumber \\
  & \times & \left. \gamma_{\nu} \frac{1}{q \sla + p \sla - k \sla_{1} 
- M + i\epsilon} \gamma_{\mu} \frac{1}{q \sla + p \sla - M +i\epsilon} 
\right\} , \\
  U_{\mu\nu}^{3} & = & {\rm tr} \left\{ \gamma_{5} \frac{1}{q \sla 
- M + i\epsilon} \gamma_{\nu} \frac{1}{q \sla + k \sla_{2} - M +i\epsilon} 
\right. \nonumber \\
  & \times & \left. \gamma_{\mu} \frac{1}{q \sla + k \sla_{1} + k \sla_{2} 
- M + i\epsilon} \gamma_{5} \frac{1}{q \sla + p \sla - M +i\epsilon} 
\right\} , \\
  U_{\mu\nu}^{4} & = & U_{\nu\mu}^{1}(k_{1} \leftrightarrow k_{2}), \\
  U_{\mu\nu}^{5} & = & U_{\nu\mu}^{2}(k_{1} \leftrightarrow k_{2}), \\
  U_{\mu\nu}^{6} & = & U_{\nu\mu}^{3}(k_{1} \leftrightarrow k_{2}). 
\label{umn2}
\end{eqnarray}
Here $M$ is the constituent u,d-quark mass. 
Because the loop integration in (\ref{epgg1}) is not divergent, 
we again do not use 
the UV cutoff. Then the gauge invariance is preserved. The inclusion of the
cutoff that is consistent with the gap equation will break the gauge 
invariance and make the present calculation too complicated.
Note that the strange quark does not contribute to the loop.
\par
   On the other hand the amplitude $T_{\mu\nu}$ has a general form required 
by the gauge invariance \cite{Eck88}
\begin{eqnarray}
  T^{\mu\nu} & = & A(x_{1},x_{2})(k_{1}^{\nu} k_{2}^{\mu} - k_{1} \cdot k_{2} 
g^{\mu\nu}) \nonumber \\
  & + & B(x_{1},x_{2}) \left[ -m_{\eta}^{2} x_{1}x_{2} g^{\mu\nu} -
\frac{k_{1} \cdot k_{2}}{m_{\eta}^{2}} p^{\mu}p^{\nu} + x_{1} k_{2}^{\mu} 
p^{\nu} + x_{2} p^{\mu} k_{1}^{\nu} \right],
\end{eqnarray}
with $x_{i}=p \cdot k_{i}/m_{\eta}^{2}$.
With $A$ and $B$, 
the differential decay rate with respect to the energies of the two photons 
is given by
\begin{eqnarray}
  \frac{d^{2} \Gamma}{dx_{1} dx_{2}} & = & \frac{m_{\eta}^{5}}{256 \pi^{2}} 
\left\{ \left|  A  + \frac{1}{2} B \right|^{2} \left[ 2(x_{1}+x_{2})
+\frac{m_{\pi}^{2}}{m_{\eta}^{2}} -1 \right]^{2} \right. \nonumber \\
  &  + &  \left. \frac{1}{4} \left| B \right| ^{2} \left[ 4 x_{1} x_{2} 
- \left[ 2(x_{1}+x_{2})+ \frac{m_{\pi}^{2}}{m_{\eta}^{2}}-1 \right] 
\right] ^{2} \right\}  \, . \label{gx1x2}
\end{eqnarray} 
Though the mass of $\eta$ as a $\bar{q} q$ bound state  depends on 
$G_D$,
we use the experimental value $m_{\eta}=547$ MeV in evaluating (\ref{gx1x2}).
The Dalitz boundary is given by two conditions:
\be
  \frac{1}{2} \left( 1-\frac{m_{\pi}^{2}}{m_{\eta}^{2}} \right) \leq  x_{1}
+x_{2} \leq 1- \frac{m_{\pi}}{m_{\eta}} ,
\ee
and
\be
  x_{1}+x_{2}-2 x_{1} x_{2} \leq \frac{1}{2} 
\left(1-\frac{m_{\pi}^{2}}{m_{\eta}^{2}} \right) .
\ee
In evaluating  (\ref{umn1})-(\ref{umn2}), one only has to identify the 
coefficients of 
$p^{\mu} p^{\nu}$ and $g^{\mu\nu}$. Details of the calculation are given 
in \cite{Ng93}. Defining ${\cal A}$ and ${\cal B}$ by
\be
  \int \frac{d^{4}q}{(2\pi)^{4}} \sum_{i=1}^{6} U_{i}^{\mu\nu} 
= -i\left( {\cal A} g^{\mu\nu} + {\cal B} \frac{p^{\mu} p^{\nu}}{m_{\eta}^{2}}
 + \cdots \right) ,
\ee
we find $A$ and $B$ as
\begin{eqnarray}
  A & = & \frac{1}{\sqrt{3}} (\cos \theta - \sqrt{2} \sin \theta) e^{2} 
g_{\pi} g_{\eta} \frac{2}{m_{\eta}^{2} \sigma} 
\left[ {\cal A}- 2 x_{1} x_{2} \
\frac{{\cal B}}{\sigma} \right] , \\
  B & = & \frac{1}{\sqrt{3}} (\cos \theta - \sqrt{2} \sin \theta) e^{2} 
g_{\pi} g_{\eta} \frac{2}{m_{\eta}^{2}} \frac{{\cal B}}{\sigma} ,
\end{eqnarray}
with
\be
  \sigma = \frac{(k_{1}+k_{2})^{2}}{m_{\eta}^{2}} = 2(x_{1}+x_{2})
+\frac{m_{\pi}^{2}}{m_{\eta}^{2}} -1 .
\ee
We evaluate  ${\cal A}$ and ${\cal B}$ numerically and further integrate 
(\ref{gx1x2}) to obtain the \epg decay rate. 
\section{Numerical results}
\subsection{$\eta$-meson mass, mixing angle and decay constant}
\hspace*{\parindent}We discuss our numerical results of the $\eta$-meson mass 
$m_\eta$, mixing angle $\theta$ and the $\eta$ decay constant $f_\eta$ in this 
subsection.  The parameters of the NJL model are the current quark masses 
$m_{u}=m_{d}$, $m_{s}$, the four-quark coupling constant $G_{S}$, the 
$U_A(1)$ breaking six-quark determinant coupling constant $G_{D}$ and the 
covariant cutoff $\Lambda$.  We take $G_{D}$ as a free parameter and study 
$\eta$ meson properties as functions of $G_{D}$.
We use the light current quark masses $m_{u}=m_{d}=8.0$ MeV to reproduce 
$M_u = M_d \simeq 330$ MeV ($\simeq 1/3 M_N$) which is the value
usually used in the nonrelativistic quark model.
Other parameters, $m_{s}$, $G_{S}$ and $\Lambda$, are determined
so as to reproduce the isospin averaged observed masses, $m_{\pi} = 138.0$ 
MeV, $m_{K} = 495.7$ MeV and the pion decay constant $f_{\pi} = 92.4$ MeV.
\par
    We obtain $m_{s}=193$ MeV, $\Lambda=783$ MeV, $M_{u,d}=325$ MeV and 
$g_{\pi q q}=3.44$, which are almost independent of $G_{D}$.
The ratio of the current s-quark mass to the current u,d-quark mass is 
$m_s/m_u = 24.1$, which agrees well with $m_s/\hat{m} = 24.4 \pm 1.5$ 
 ($\hat{m} = \frac{1}{2} (m_u + m_d)$) derived from ChPT \cite{L96}.
The kaon decay constant $f_K$ is the prediction and is almost independent of 
$G_D$.  We have obtained $f_K = 97$ MeV which is about 
14\% smaller than the observed value.  We consider this is the typical 
predictive power of the NJL model in the strangeness sector. 
The quark condensates are also independent of $G_D$ and our results are 
$\langle \bar uu \rangle^{\frac{1}{3}} = -216$ MeV and 
$\langle \bar ss \rangle^{\frac{1}{3}} = -226$ MeV, which give 
\be
\frac{f_\pi^2 m_\pi^2}{-2 m_u \langle \bar uu \rangle} = 1.01 ,\quad \quad
\frac{f_K^2 m_K^2}{-\frac{1}{2}(m_u + m_s) \langle \bar uu + \bar ss \rangle}
= 1.06 \, .
\ee
As reported in the previous studies in the NJL model, the 
Gell-Mann-Oakes-Renner relations hold well for both the pion and kaon sectors.
\par
\begin{figure}[t]
\begin{center}
\leavevmode
\epsfxsize=300pt
\epsfbox{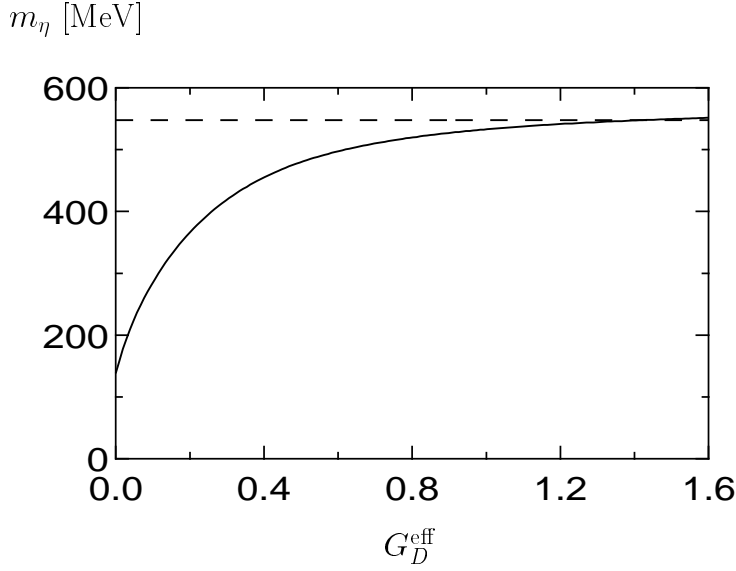}    
\caption{Dependence of the $\eta$ meson mass
on the dimension-less coupling constant $G_{D}^{\rm eff}$. The horizontal
dashed line indicates the experimental value.\hfil\break}
\label{fig:mass}
\end{center}
\end{figure}
\begin{figure}[t]
\begin{center}
\leavevmode
\epsfxsize=300pt
\epsfbox{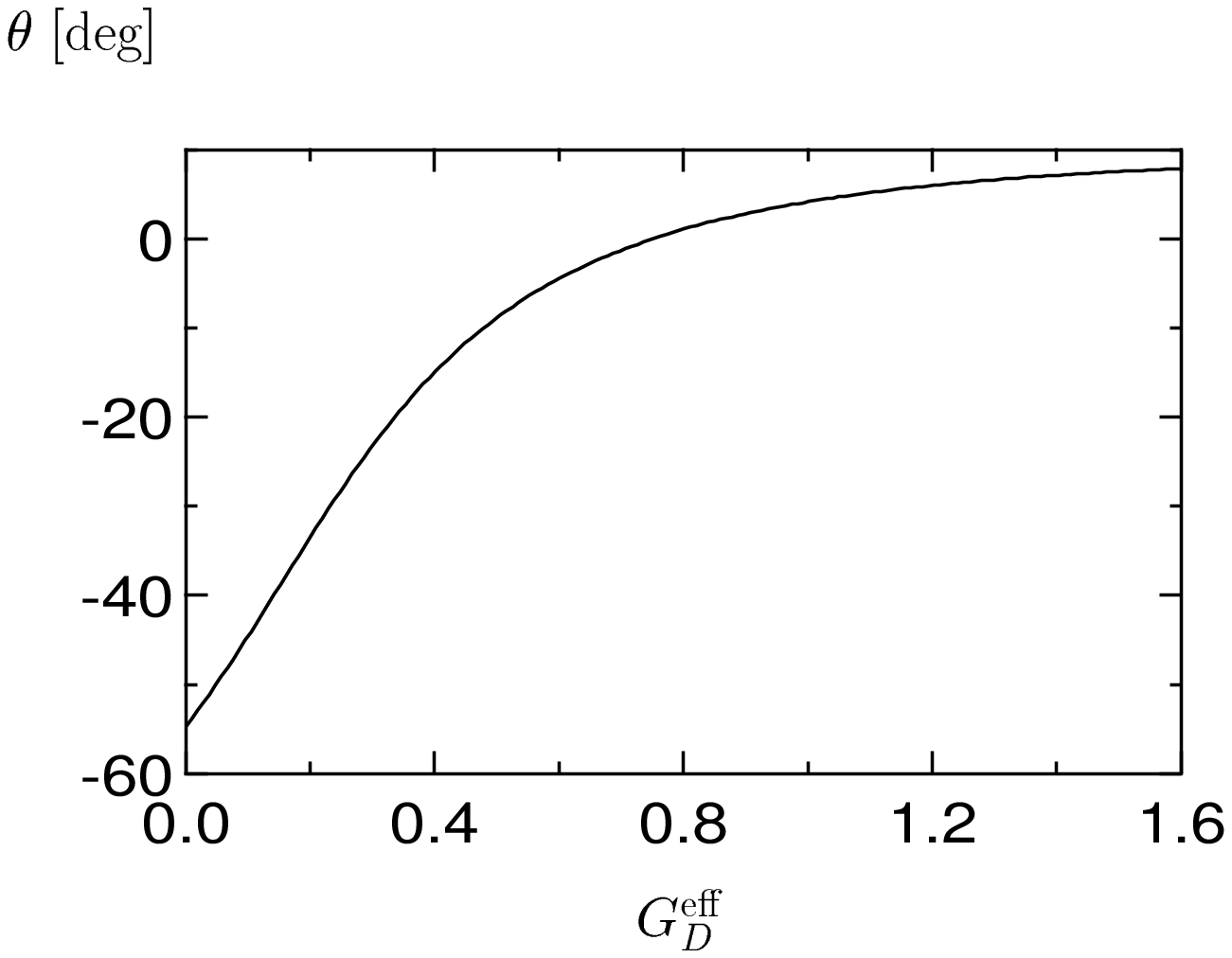}    
\caption{Dependence of the mixing angle $\theta$
on the dimension-less coupling constant $G_{D}^{\rm eff}$. The horizontal
dashed line indicates the experimental value.\hfil\break}
\label{fig:mix}
\end{center}
\end{figure}

    We define dimensionless parameters 
$G_{D}^{\rm eff} \equiv - G_{D} (\Lambda / 2 \pi)^{4} \Lambda N_{c}^{2}$ and 
$G_{S}^{\rm eff} \equiv G_{S} (\Lambda / 2 \pi)^{2} N_{c}$. 
The calculated results of the $\eta$-meson mass $m_\eta$ and the mixing 
angle $\theta$ are shown in Fig. \ref{fig:mass} and Fig. \ref{fig:mix}, 
respectively.
When $G_{D}^{\rm eff}$ is zero, our lagrangian does not cause the flavor 
mixing and therefore the ideal mixing is achieved. 
The ``$\eta$'' is purely $u\bar u + d\bar d$ which corresponds to 
$\theta = -54.7^{\circ}$ and is degenerate to the pion in this limit.
\par
\begin{figure}[t]
\begin{center}
\leavevmode
\epsfxsize=300pt
\epsfbox{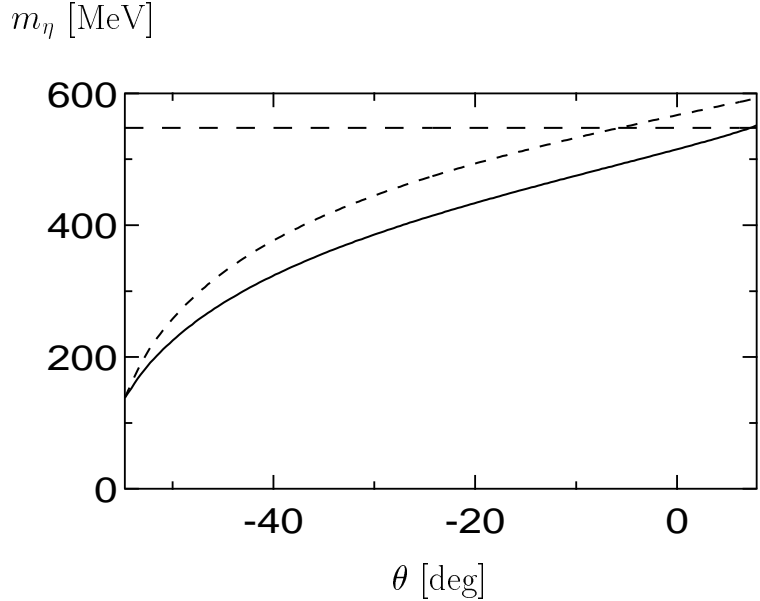}    
\caption{Dependence of the $\eta$ meson mass on the mixing angle.  The solid 
line indicates the result calculated in the NJL model and the short-dashed 
line indicates that in the $1/N_C$ expansion approach.  The horizontal 
long-dashed line shows the experimental $\eta$ meson mass. \hfill\break}
\label{fig:massnc}
\end{center}
\end{figure}

    It may be useful to compare our results with those in the $1/N_C$ 
expansion approach.  
In the $1/N_C$ expansion approach, with the inclusion of the 
$O(1/N_C)$ contribution by the $U_A(1)$ anomaly in the flavor singlet-singlet 
channel, the square mass matrix of the mass term of the low-energy effective 
lagrangian in the $\eta_8$-$\eta_0$ channel becomes as follows \cite{Vene79}. 
\be
M_{\eta - \eta'}^2 = \left(
\begin{array}{cc}
\frac{4}{3} m_K^2 - \frac{1}{3} m_\pi^2 & 
- \frac{2}{3} \sqrt{2} ( m_K^2 - m_\pi^2 ) \\
- \frac{2}{3} \sqrt{2} ( m_K^2 - m_\pi^2 ) & 
\frac{2}{3} m_K^2 + \frac{1}{3} m_\pi^2 + \frac{\chi^2}{N_C} 
 \end{array}
\right) \, , \label{etaetpmm}
\ee
where
\be
\frac{\chi^2}{N_C} \equiv \frac{6}{f_\pi^2} (-i) \int d^4 x \partial^\mu 
\partial^\nu T \langle K_\mu (x) K_\nu (0) \rangle_{YM} \, .
\ee
Here the isospin symmetry is assumed and YM means the pure Yang-Mills theory.
The ghost field $K_\mu$ is defined 
by
\be
\partial^\mu K_\mu \equiv \frac{g^2}{32 \pi^2} 
G_{\mu \nu}^a (\widetilde G^a)^{\mu \nu} \, , 
\ee
and $G_{\mu \nu}^a$ is the gluon field strength tensor.
By diagonizing the matrix given in Eq. (\ref{etaetpmm}), we obtain 
\be
m_{\eta, \eta'}^2 = \left( m_K^2 + \frac{\chi^2}{2N_C} \right) \pm \frac{1}{2}
\sqrt{\left( 2 m_K^2 - 2 m_\pi^2 - \frac{\chi^2}{3 N_C} \right)^2 
             + \frac{8}{9} \frac{\chi^4}{N_C^2} } \, , \label{etaetpmass}
\ee
and 
\be
\tan \theta = \frac{\frac{4}{3} m_K^2 - \frac{1}{3} m_\pi^2 - m_\eta^2}{
- \frac{2}{3} \sqrt{2} (m_K^2 - m_\pi^2)} \, . \label{etaetpmix}
\ee
From Eqs. (\ref{etaetpmass}) and (\ref{etaetpmix}), it is obvious that in 
the $U_A(1)$ limit the $\eta$ and $\eta'$ become the ideal mixing state with 
$m_\eta = m_\pi \simeq 138$ MeV and $m_{\eta'} = \sqrt{2 m_K^2 - m_\pi^2} 
\simeq 687$ MeV.  
We compare the dependence of the $\eta$ meson mass on the mixing angle 
calculated in the NJL model with that given in Eq. (\ref{etaetpmix}) in the 
$1/N_C$ expansion approach in Fig. \ref{fig:massnc} and find that the $\eta$ 
meson mass calculated in the NJL model is somewhat smaller than that in the 
$1/N_C$ expansion approach at the same mixing angle (except for the ideal 
mixing point) though the shapes are similar.  
The value of the mixing angle is one of the important quantity to understand
the physics of the $\eta$ and $\eta'$ mesons.  The mixing angle determined 
from the $\eta$-$\eta'$ mass formula is often discussed in the literature.
However since the $\eta$ meson mass is not so sensitive 
to the mixing angle as shown in Fig. \ref{fig:massnc}, it is not suitable to
determine the mixing angle from the $\eta$ meson mass.
\par
\begin{figure}[t]
\begin{center}
\leavevmode
\epsfxsize=280pt
\epsfbox{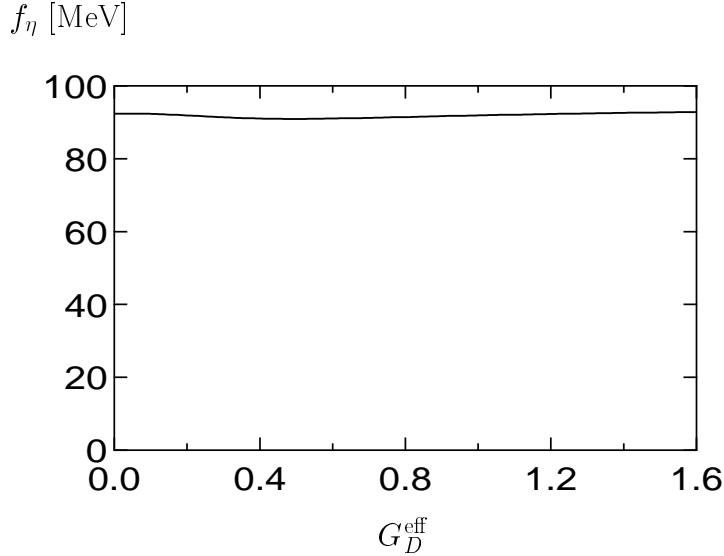}    
\caption{Dependence of the $\eta$ decay constant $f_\eta$
on the dimension-less coupling constant $G_{D}^{\rm eff}$.\hfil\break}
\label{fig:dc}
\end{center}
\end{figure}

    We next discuss the $\eta$ decay constant $f_\eta$.  The calculated $\eta$
decay constant is shown in Fig. \ref{fig:dc}.  It is almost independent of 
$G_D$ and $f_\eta \simeq f_\pi$.  For example, $f_\eta =91.2$ MeV at 
$G_D^{\rm eff} = 0.7$.  Therefore it seems that the $\eta$ meson does not lose
the Nambu-Goldstone boson nature though its mass and mixing angle are strongly 
affected by the $U_A(1)$ breaking interaction.
\par
    The charged pion and kaon decay constants can be directly obtained by 
measuring the $\pi \to \mu \nu_\mu$ and $K \to \mu \nu_\mu$ decays.  
On the other hand the decay constants for the light neutral pseudoscalar 
mesons $\pi^0$, $\eta$ and $\eta'$ cannot be obtained from the direct 
measurements.  As discusses in Sect. 3.1, they are related to the 
$P \to \gamma \gamma$ decay amplitudes in the soft limit.  
It is widely accepted that the soft meson limit is close to the on-shell point 
in the pion case.  However it is the matter of discussions how good the 
soft $\eta$ and $\eta'$ limits are.  We will discuss this point in the next 
section by comparing the \egg decay amplitude which is directly calculated 
in the NJL model with that obtained using PCAC + ABJ anomaly in the soft 
$\eta$ limit.  
\par
    Since the NJL model does not confine quarks, the $\eta'$ meson is not a 
bound state.  Therefore we do not apply our model to the $\eta'$ meson.
On the contrary, the $\eta$ meson appears as a bound state.  
Nevertheless, one may ask whether the present model is applicable to the 
$\eta$ meson since the binding energy of the $\eta$ meson ($2 M_u - m_\eta$) 
is much smaller than those of the pion and kaon. In order to confirm it, we 
study the constituent u, d-quark mass dependence of the $\eta$-meson 
properties.  By changing the current u, d-quark mass $m_{u,d}$ from 7.5 MeV to 
8.5 Mev, the constituent u,d-quark mass $M_{u,d}$ is changed from about 
300 MeV to 360 MeV.  Other parameters of the model are chosen so as to 
reproduce the experimental values of $m_\pi$, $m_K$, $m_\eta$ and $f_\pi$. 
This change causes the change of the binding energy of the $\eta$ meson from
about 50 MeV to 170 MeV. The change of the calculated $\eta$ decay constant 
is within 2\% and the change of the calculated mixing angle is within 
12\%.  This stability of the $\eta$-meson properties indicates that the NJL 
model can describe the essential feature of the $\eta$ meson.  
\subsection{\egg decay}
\hspace*{\parindent}The recent experimental results of the 
$\pi^0, \eta \to \gamma \gamma$ decay
 widths are 
$\Gamma(\pi^0 \to \gamma \gamma) = 7.7 \pm 0.6 \, {\rm eV}$ and 
$\Gamma(\eta \to \gamma \gamma) = 0.510 \pm 0.026 \, {\rm keV}$ \cite{PDG} 
and the reduced amplitudes are 
\begin{eqnarray}
\left\vert \widetilde{\cal T}_{\pi^0 \to \gamma \gamma} \right\vert & = &
(2.5 \pm 0.1) \times 10^{-11} \, [{\rm eV}]^{-1} \, , \label{exppidw} \\
\left\vert \widetilde{\cal T}_{\eta \to \gamma \gamma} \right\vert & = &
(2.5 \pm 0.06) \times 10^{-11} \, [{\rm eV}]^{-1} \, . \label{expetadw}
\end{eqnarray}
Here we have used the two photon measurement result for the \egg decay width.
{}From Eq. (\ref{amplpi}) and Eq. (\ref{ampleta}), we get 
$\widetilde{\cal T}_{\eta \to \gamma \gamma} = (5/3) 
\widetilde{\cal T}_{\pi^0 \to \gamma \gamma}$ in the $U_A(1)$ limit.  
Therefore in order to reproduce the experimental value of 
$\widetilde{\cal T}_{\eta \to \gamma \gamma}$, the effect of the 
$U_A(1)$ anomaly should reduce 
$\widetilde{\cal T}_{\eta \to \gamma \gamma}$ by a factor 3/5.
\par
    We first discuss the $\pi^0 \to \gamma \gamma$ decay.  
The calculated result is 
$\widetilde{\cal T}_{\pi^0 \to \gamma \gamma} = 
2.50 \times 10^{-11} (1/{\rm eV})$
which agrees well with the observed value given in Eq. (\ref{exppidw}).
The current algebra result is 
$\widetilde{\cal T}_{\pi^0 \to \gamma \gamma} = 
\alpha/(\pi f_\pi) = 2.514 \times 10^{-11} (1/{\rm eV})$, 
and thus the soft pion limit is a good approximation for 
$\pi^0 \to \gamma \gamma$ decay.  
The chiral symmetry breaking affects
$\widetilde{\cal T}_{\pi^0 \to \gamma \gamma}$ in two ways.  One is
the deviation from  
the G-T relation and another is the matrix element of the triangle diagram 
$F(u,\pi^0)$.  Our numerical results are $g_\pi = 3.44$, $M_u/f_\pi = 3.52$
and $F(u,\pi^0) M_u = 1.015$, therefore the deviations from the soft pion 
limit are very small both in the G-T relation and the matrix element of the
triangle diagram.
\par
\begin{figure}[t]
\begin{center}
\leavevmode
\epsfxsize=280pt
\epsfbox{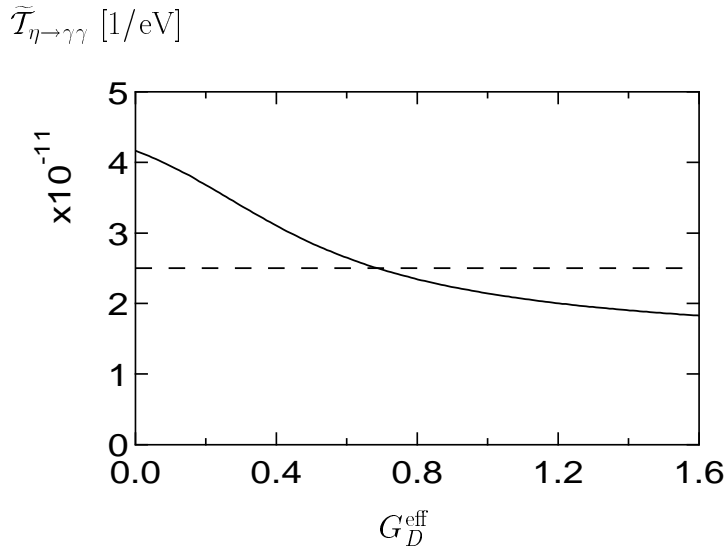}    
\caption{Dependence of the $\eta \to \gamma \gamma $ decay amplitude
on the dimension-less coupling constant $G_{D}^{\rm eff}$. The horizontal
dashed line indicates the experimental value.\hfil\break}
\label{fig:e2gg}
\end{center}
\end{figure}

    Let us now turn to the discussion of the \egg decay.  The calculated 
results of the \egg decay amplitude 
$\widetilde{\cal T}_{\eta \to \gamma \gamma}$ is given in Fig. \ref{fig:e2gg}. 
The experimental value of the \egg decay amplitude is reproduced at about 
$G_D^{\rm eff} = 0.7$. The calculated $\eta$-meson mass at 
$G_D^{\rm eff} = 0.7$ is $m_{\eta} = 510$ MeV which is 7\% smaller than 
the observed mass.  
As for the effects of the symmetry breaking on 
$\widetilde{\cal T}_{\eta \to \gamma \gamma}$, our results are 
$F(u, \eta)\, M_u = 1.36$ and $F(u, \eta)/F(s, \eta) = 1.99$ at 
$G_D^{\rm eff} = 0.7$.   Therefore it seems that the soft $\eta$ limit 
is not close to the real world in this case. 
Recently, Bernard et al. \cite{BBHMR} calculated the \egg decay width using a 
similar model.  They used a rather weak instanton induced interaction and 
their result of $\Gamma(\eta \to \gamma \gamma)$ is about 50\% bigger than 
the experimental value.  It is understandable from our analysis.
\par
    $G_D^{\rm eff} = 0.7$ corresponds to 
$G_{D} \langle \overline{s} s \rangle / G_{S} =0.44$, suggesting that 
the contribution from ${\cal L}_{6}$ to the dynamical mass of 
the up and down quarks is 44\% of that from ${\cal L}_{4}$.
In the previous study of the $\eta$ and $\eta'$ mesons in the three-flavor 
NJL model \cite{KH88,HK94,BJM88,KKT88,RA88,KLVW90}, the strength of the 
instanton induced interaction has been determined so as to reproduce the 
observed $\eta'$ mass though the $\eta'$ state has the unphysical decay 
mode of the $\eta' \to \bar uu , \bar dd$.  
The strength determined from the $\eta'$ 
is much smaller than $G_D^{\rm eff} = 0.7$, about 1/4 of the present case.
One of the shortcomings of the NJL model is the lack of the confinement 
mechanism.  It is expected that the confinement gives rise to the attractive 
force between quark and antiquark in the $\eta'$ meson to prevent the $\eta'$ 
meson from decaying to the quark and antiquark pair. 
\par
    In the PCAC + ABJ anomaly approach, if one assumes the $SU(3)$ symmetry, 
i.e., $f_8 = f_\pi$, and using Eqs. (\ref{etaggpcac}) and (\ref{etpggpcac}),   
the mixing angle $\theta$ and the meson decay constant in the flavor singlet 
channel $f_0$ can be determined so as to reproduce the observed 
$\Gamma(\eta \to \gamma \gamma)$ and $\Gamma(\eta' \to \gamma \gamma)$.
The results are $\theta = -17.4^{\circ}$ and $f_0/f_\pi = 1.1$.
The pion and kaon loop corrections are then included in ChPT \cite{DHL85} and 
the results are $\theta = -21.8^{\circ}$, $f_8/f_\pi = 1.25$ and 
$f_0/f_\pi = 1.07$.     
\par
    In the NJL model, the mixing angle at $G_{D}^{\rm eff}=0.7$ is 
$\theta = -1.3^{\circ}$ and that indicates a strong OZI violation and a 
large (u,d)-s mixing.  This disagrees with the ``standard'' value 
$\theta \simeq -20^{\circ}$ obtained in  the PCAC + ABJ anomaly approach and 
ChPT.   One of the origin of the difference is that the mixing angle 
in the NJL model depends on $p^{2}$ of the $\overline{q}q$ state 
and thus reflects the internal structure of the $\eta$ meson. 
On the contrary the analyses in the PCAC + ABJ anomaly approach and ChPT
assume an energy-independent mixing angle, i.e., 
$\theta(p^2=m_{\eta}^{2})=\theta(p^2=m_{\eta'}^{2})$.
Another point is that the $SU(3)$ breaking of the \egg decay amplitude 
is rather large in the NJL model as we have shown above. 
It indicates that the soft $\eta$ and $\eta'$ limits used in the 
PCAC + ABJ anomaly approach are not so good.  In the ChPT point of view, 
it suggests that the tree diagram contributions from $O(p^6)$ terms may 
be rather large. 
\begin{figure}[t]
\begin{center}
\leavevmode
\epsfxsize=280pt
\epsfbox{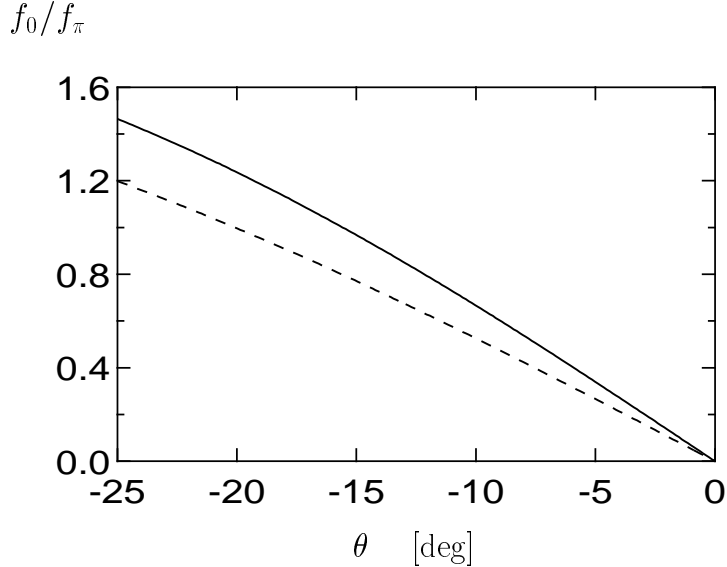}    
\caption{The flavor singlet decay constant $f_0$ which is determined so as 
to reproduce the observed \egg decay width as a function of $\theta$ using the 
formula derived in the PCAC + ABJ anomaly approach.  The solid line indicates 
in the case of $f_8 = f_\pi$ and the dashed line indicates in the case 
of $f_8 = 1.25 \, f_\pi$.}
\label{fig:dc0}
\end{center}
\end{figure}
\par
    If the PCAC + ABJ anomaly approach is considered not to be applied for 
the $\eta' \to \gamma \gamma$ decay, then one cannot determine the mixing 
angle $\theta$ and the flavor singlet decay constant $f_0$ from the observed 
$\eta, \eta' \to \gamma \gamma$ decay widths.  
However the relation between $\theta$ and $f_0$ can be obtained from 
Eq. (\ref{etaggpcac}) only by the \egg decay amplitudes assuming that 
$f_8$ is given.  Fig. \ref{fig:dc0} shows the $f_0 - \theta$ relations for 
the $SU(3)$ value $f_8 = f_\pi$ and the ChPT estimation $f_8 = 1.25 f_\pi$.
One sees that a smaller flavor singlet component of the $\eta$ meson 
corresponds to a smaller $f_0$. 
It is not strange that $f_0$ is smaller than $f_\pi$ since the $\eta'$ 
meson has larger gluonic component than pion because of the $U_A(1)$ 
anomaly. 
\subsection{\egl decay}
\begin{figure}[t]
\begin{center}
\leavevmode
\epsfxsize=300pt
\epsfbox{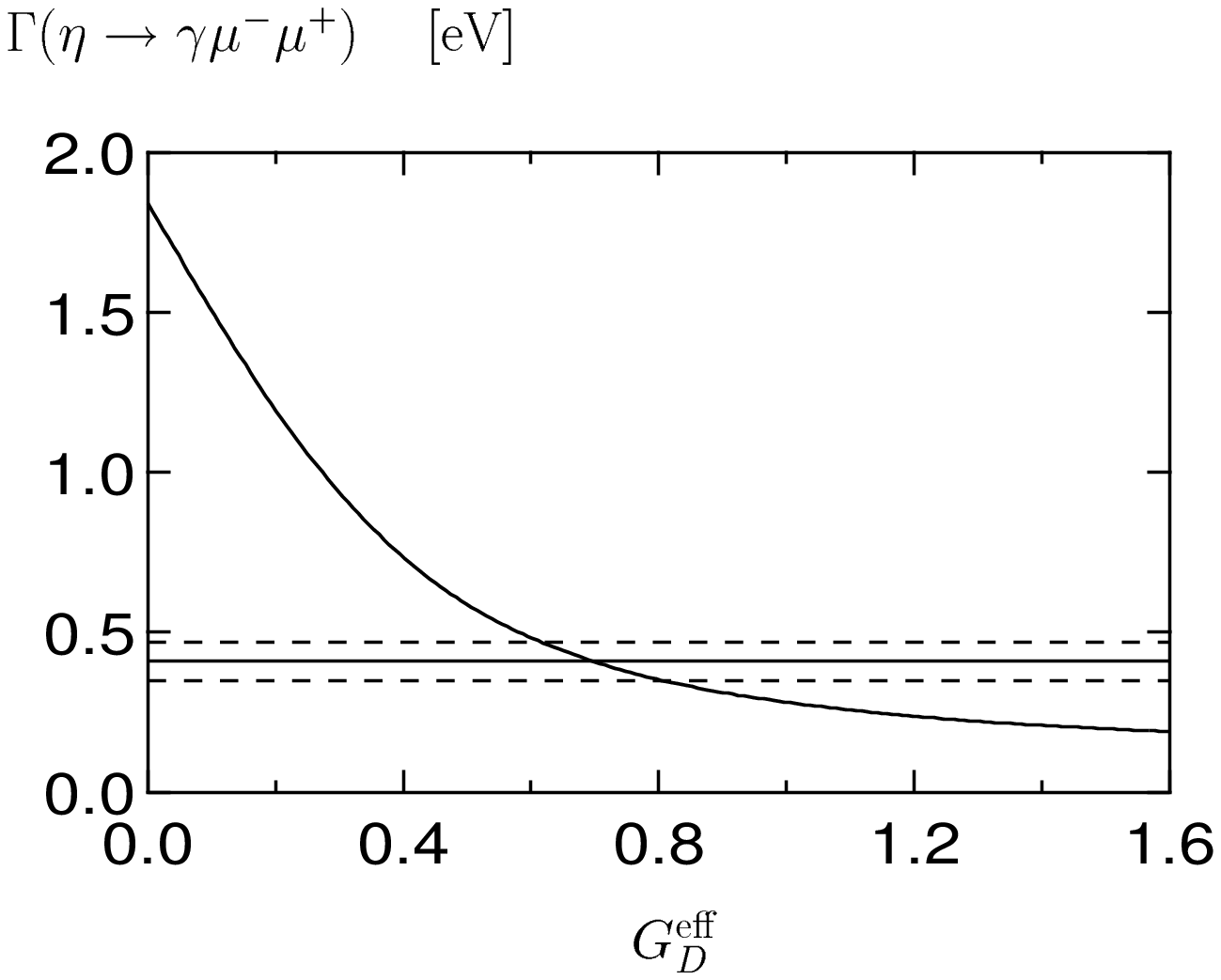}    
\caption{Dependence of the \egm decay width on the dimension-less coupling 
constant $G_{D}^{\rm eff}$. The horizontal solid line indicates the 
experimental value and the dashed lines indicate its error widths.}
\label{fig:e2gmm}
\end{center}
\end{figure}

\begin{figure}[t]
\begin{center}
\leavevmode
\epsfxsize=300pt
\epsfbox{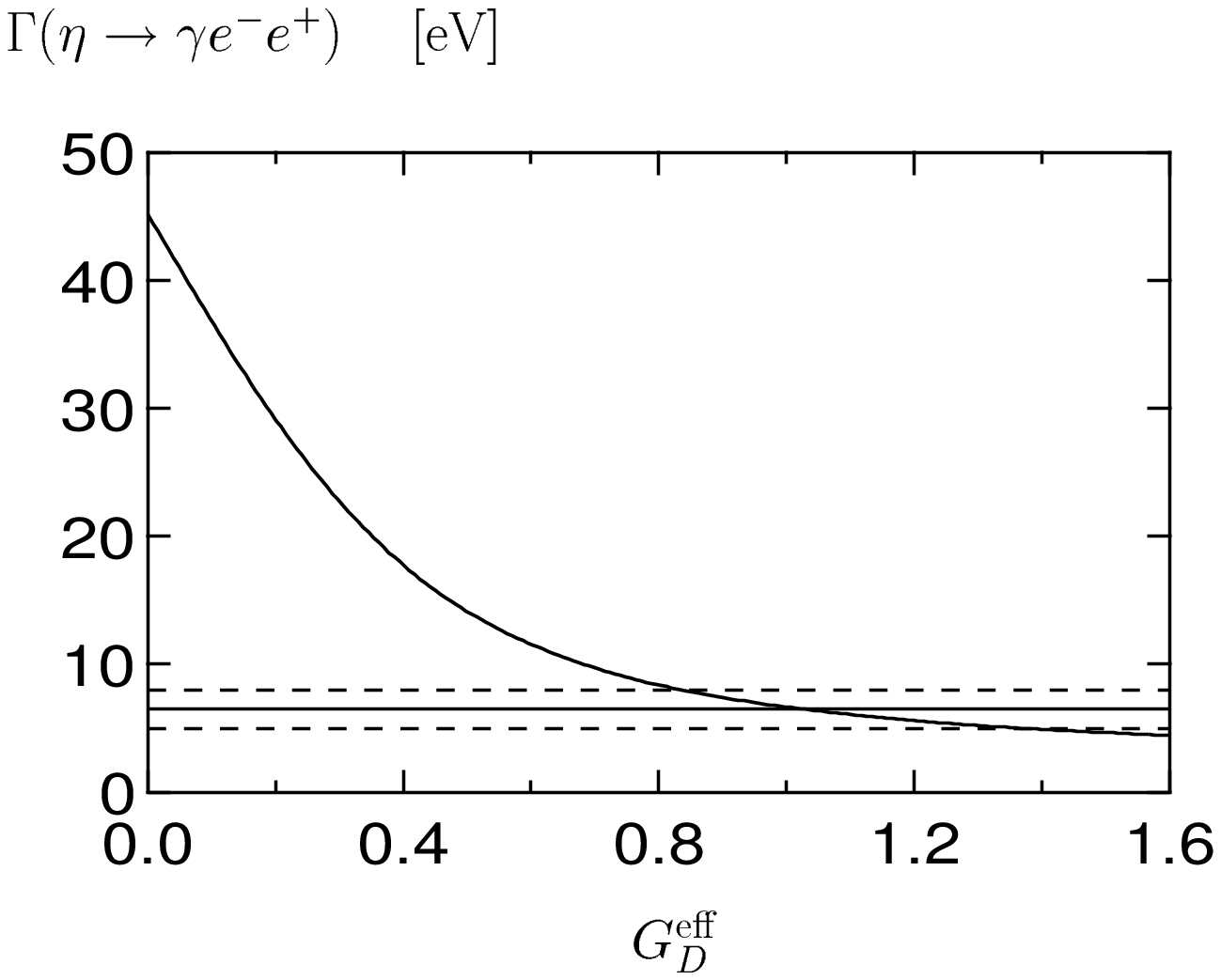}    
\caption{Dependence of the \ege decay width on the dimension-less coupling 
constant $G_{D}^{\rm eff}$. The horizontal solid line indicates the 
experimental value and the dashed lines indicate its error widths.}
\label{fig:e2gee}
\end{center}
\end{figure}

\hspace*{\parindent}The experimental value of the \egm and \ege decay widths 
are \cite{PDG} \footnote{We have used the two photon measurement result of 
the \egg decay width: $\Gamma(\eta \to \gamma \gamma) = 0.510 \pm 0.026$keV.}
\bea
\Gamma \left( \eta \to \gamma \mu^- \mu^+ \right) & = & 
0.41 \pm 0.06 {\rm \ eV} \, , \label{expegmdw} \\
\Gamma \left( \eta \to \gamma e^- e^+ \right) & = & 
6.5 \pm 1.5 {\rm \ eV} \, , \label{expegedw}
\eea
and the calculated \egm and \ege decay widths are shown in 
Fig. \ref{fig:e2gmm} and Fig. \ref{fig:e2gee} respectively.
At $G_{D}^{\rm eff}=0.70$ where the \egg decay width is reproduced, we obtain 
$\Gamma \left( \eta \to \gamma \mu^- \mu^+ \right) = 0.407$ eV, which is in 
good agreement with the experimental result shown in (\ref{expegmdw}). 
On the other hand, our calculated result of the \ege at $G_{D}^{\rm eff}=0.70$
is $\Gamma \left( \eta \to \gamma e^- e^+ \right) = 9.72$ eV, which is about 
50\% bigger than the observed value given in (\ref{expegedw}).  
\par
    It is clear from the electron-positron pair invariant mass square spectrum 
of the $\eta \to \gamma e^- e^+$ decay given in Eq. (\ref{p2gllsp}) that 
the strength is mostly saturated in the small region just above 
$q^2 = (2 m_e)^2$.  Since in this region, the $\eta \gamma \gamma^*$ 
transition form factor $f_{\eta \gamma \gamma^*}(q^2)$ is almost unity and the
radiative corrections are found to be negligible \cite{LS71}, the $\eta \to 
\gamma e^- e^+$ decay width is strongly related to the \egg decay width and 
it is rather hard to explain the present experimental values of 
$\Gamma (\eta \to \gamma \gamma)$ and $\Gamma (\eta \to \gamma e^- e^+)$ 
simultaneously.
\par
\begin{figure}[t]
\begin{center}
\leavevmode
\epsfxsize=300pt
\epsfbox{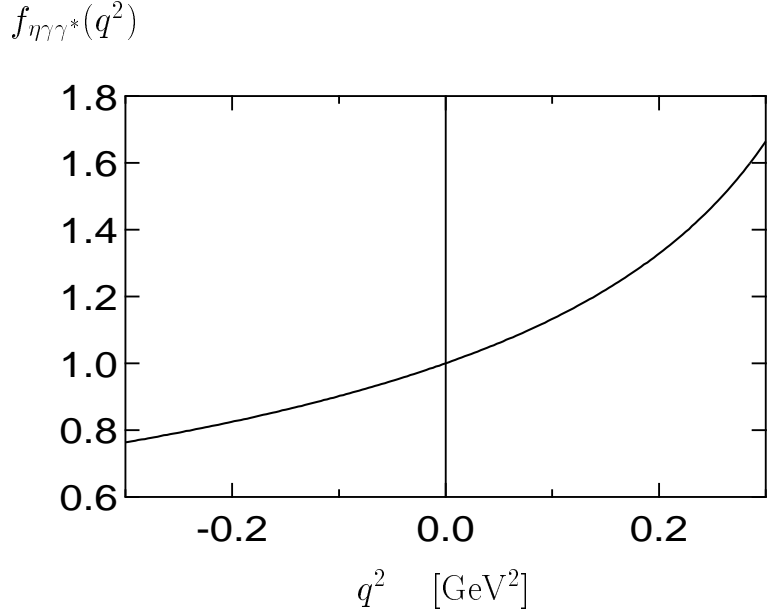}    
\caption{$q^2$-dependence of the $\eta \gamma \gamma^*$ transition form factor 
$f_{\eta \gamma \gamma^*}(q^2)$ calculated at $G_D^{\rm eff} = 0.7$} 
\label{fig:e2gllff}
\end{center}
\end{figure}

    The calculated $\eta \gamma \gamma^*$ transition form factor 
$f_{\eta \gamma \gamma^*}(q^2)$ at $G_D^{\rm eff} = 0.7$ is shown in 
Fig. \ref{fig:e2gllff} and the calculated slope parameter defined in 
Eq. (\ref{lamdap}) is $\Lambda_{\eta} = 0.94$ GeV.
The recent experimental results of the slope parameter are as follows. 
The ${\rm TPC}/2\gamma$ Collaboration at the SLAC gives 
$\Lambda_{\eta} = 0.70 \pm 0.08$ GeV \cite{TPC} and the CELLO Collaboration 
at the DESY gives $\Lambda_{\eta} = 0.84 \pm 0.06$ GeV \cite{CELLO}. 
So our result is somewhat larger than the experimental results.
As for the slope parameter of the $\pi^0 \gamma \gamma^*$ form factor,
our model gives the simple result in the chiral limit, i.e., 
$\Lambda_{\pi} = \sqrt{12} M_u \simeq 1.126$ GeV, which is also larger than 
the CELLO result $\Lambda_{\pi} = 0.75 \pm 0.03$ GeV \cite{CELLO}.
\par
    Ametller et al. \cite{Ame92-1} studied the transition form factors for the 
$P \gamma \gamma^*$ vertices with $P = \pi^0$, $\eta$ and $\eta'$ using the 
most successful and/or traditional models of the low-energy QCD including the 
vector meson dominance model (VMD), the constituent quark loop model (QL), 
the QCD-inspired interpolation model by Brodsky-Lepage (BL) and the ChPT.
They concluded that all the models considered agree in the correct value for 
a mean $\Lambda_P$, but differ in the breaking pattern of the $SU(3)$ flavor 
symmetry.  Our approach is close to QL approach. In Ref.  \cite{Ame92-1},
a rather small constituent quark mass ($M_{u,d} \simeq 0.23$ GeV) has 
been used and that is the main reason why QL prediction of $\Lambda_{\eta}$
is in reasonable agreement with the experimental results. However, it should 
be noted here that since the constituent quark mass used in QL is smaller 
than $m_{\eta}/2$, the unphysical imaginary part appears in the 
$\eta \to \gamma \gamma^*$ transition amplitude corresponding to the 
unphysical channel $\eta \to \bar uu ,\, \bar dd$.
\par
    Our interpretation of the present result in the NJL model is as follows.
Since the pseudoscalar mesons in the NJL model have the quark-antiquark 
structures and therefore have the size.  For the $\pi^0 \gamma \gamma^*$
vertices, it is $r_{\pi^0} \simeq 1/\sqrt{2} M_u \simeq 0.43$ fm, which 
is about 67\% of the experimental value 
$r_{\pi^0} = \sqrt{6}/\Lambda_\pi \simeq 0.64$ fm.  
Our result of the $\eta \gamma \gamma^*$ vertex size is 
$r_{\eta} = \sqrt{6}/\Lambda_\eta \simeq 0.51$ fm, which should be compared 
with the experimental results $r_{\eta} \simeq 0.69$ fm \cite{TPC} and 
$r_{\eta} \simeq 0.58$ fm \cite{CELLO}.  
The introduction of the quark-antiquark correlations in the 
vector channel may solve the problem of the difference of the sizes for the 
$P \gamma \gamma^*$ vertices between our results and the experimental ones.
\subsection{\epg decay}
\begin{figure}[t]
\begin{center}
\leavevmode
\epsfxsize=300pt
\epsfbox{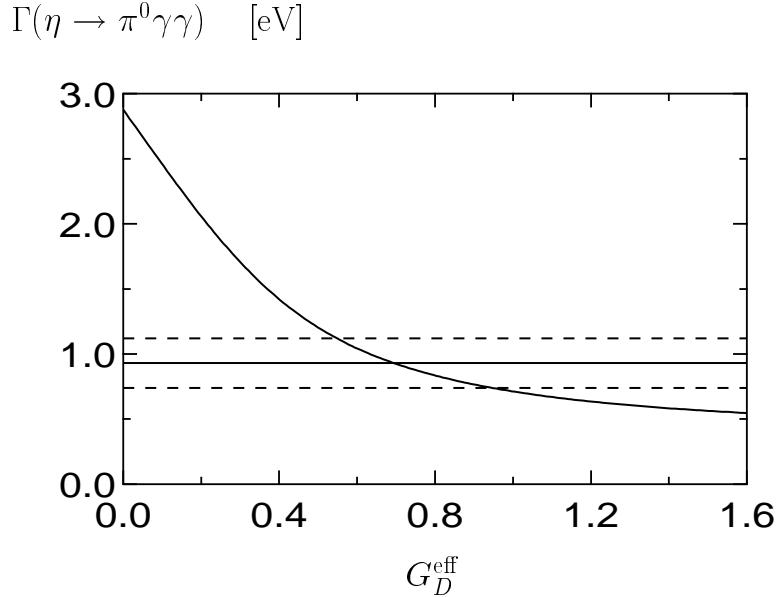}    
\caption{Dependence of the \epg decay width on the dimension-less coupling 
constant $G_{D}^{\rm eff}$. The horizontal solid line indicates the 
experimental value and the dashed lines indicate its error widths.}
\label{fig:e2pgg}
\end{center}
\end{figure}

\hspace*{\parindent}The experimental value of the \epg decay width 
is \cite{PDG} \footnote{We have used the two photon measurement result of 
the \egg decay width: $\Gamma(\eta \to \gamma \gamma) = 0.510 \pm 0.026$keV.}
\be
  \Gamma \left( \eta \rightarrow \pi^{0} \gamma \gamma \right) 
= 0.93 \pm 0.19 {\rm \ eV} \, , \label{expepdw}
\ee
and the calculated \epg decay width is shown in Fig. \ref{fig:e2pgg}.
At $G_{D}^{\rm eff}=0.70$ where the \egg decay width is reproduced, we obtain 
$\Gamma(\eta \rightarrow \pi^{0} \gamma \gamma )=0.92$eV, which is in good 
agreement with the experimental data shown in (\ref{expepdw}). 
\par
    In ChPT \cite{Ame,Jett}, since the lowest order $O(p^2)$ term and the 
next order $O(p^4)$ tree diagrams do not contribute to the \epg process, 
the $O(p^4)$ one-loop diagrams are the leading contributions.  The calculated 
pion and kaon one-loop contributions to the \epg decay width  are 
$\Gamma^{(4)}_\pi (\eta \to \pi^0 \gamma \gamma) = 
0.84 \times 10^{-3}{\rm eV}$, 
$\Gamma^{(4)}_K (\eta \to \pi^0 \gamma \gamma) = 
2.45 \times 10^{-3}{\rm eV}$ and 
$\Gamma^{(4)}_{\pi, K} (\eta \to \pi^0 \gamma \gamma) = 
0.84 \times 10^{-3}{\rm eV}$,
which are more than two orders of magnitude below the observed width
(\ref{expepdw}).
This is because the pion loop violates the G-parity 
invariance and the kaon loop is also suppressed by the large kaon mass.
At $O(p^{6})$, there exists contribution coming from tree diagrams, 
one-loops and two-loops. The loop contributions are smaller than those 
from the order $O(p^{4})$. 
In \cite{Ame}, the coupling strengths of the $O(p^6)$ tree diagrams are 
determined assuming saturation by the scalar meson $a_0(980)$ and the 
tensor meson $a_2(1310)$ resonances as well as the $\rho$ and $\omega$ 
vector meson resonances.  The obtained resonance contributions are 
$\Gamma^{(6)}_{\rho + \omega}(\eta \to \pi^0 \gamma \gamma) = 0.18 {\rm eV}$
and $\Gamma^{(6)}_{\rho + \omega + a_0 + a_2}(\eta \to \pi^0 \gamma \gamma) 
= 0.18 \pm 0.02 {\rm eV}$.  
The contributions of other mesons,  such as 
$b_1(1235),h_1(1170),h_1(1380)$ \cite{Ko2} 
and other tree diagrams \cite{Ko}, are found to be small.
Therefore, up to $O(p^6)$, ChPT estimation of the \epg decay width is 
about 1/4 of the experimental result.  
\par
On the other hand in \cite{Bel} the $O(p^{6})$ tree diagrams are evaluated by 
using the extended NJL (ENJL) model\cite{Bijnens}. They calculated three
contributions in ENJL, namely, the vector and scalar resonance exchange and
the quark-loop contributions. Their result is $\Gamma(\eta \to \pi^0 \gamma
\gamma) \simeq 0.5$ eV. They further introduced the $O(p^{8})$ chiral 
corrections as well as the axialvector and tensor meson exchange contributions,
and finally obtained $\Gamma(\eta \to \pi^0 \gamma \gamma) = 0.58 \pm 0.3$ eV.
A recent similar analysis in ENJL concludes somewhat smaller value 
for the decay width \cite{Bijnens3}.
The difference between our approach and that in \cite{Bel} are as follows.
The ENJL model lagrangian has not only the scalar-pseudoscalar four quark 
interactions but also the vector-axialvector four quark interactions.
However, the $U_{A}(1)$ breaking is not explicitly included in their model
and therefore the $\eta-\eta'$ mixing is introduced by hand with the mixing
angle $\theta = -20^{\circ}$. We stress that the introduction of the $U_{A}(1)$
breaking interaction is important to understand the structure of the $\eta$
meson. 
There is another difference. The coupling constants of the chiral effective
meson lagrangian predicted in the ENJL model are parameters of the Green
function evaluated at zero momenta. On the other hand we evaluate the 
quantities at the pole position of the mesons.
\begin{figure}[t]
\begin{center}
\leavevmode
\epsfxsize=300pt
\epsfbox{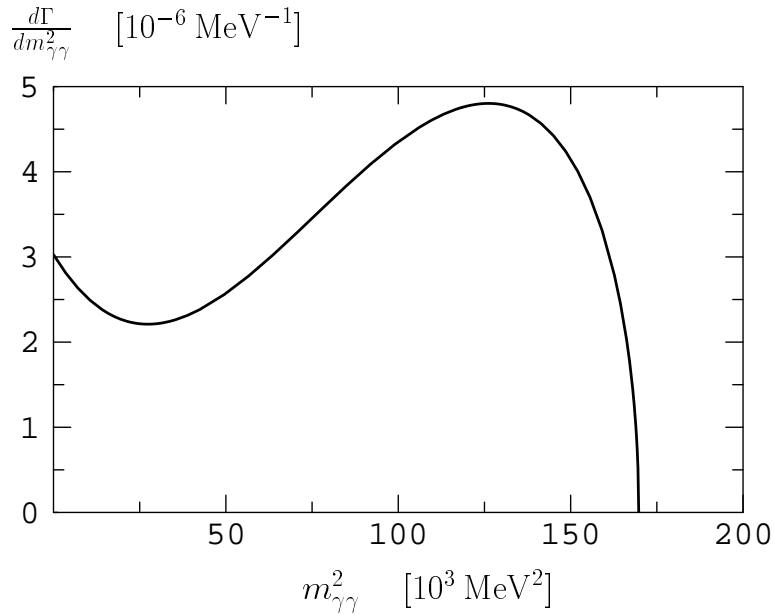}    
\caption{Spectrum of the photon invariant mass $m_{\gamma\gamma}^{2}$.}
\label{fig:e2pggsp1}
\end{center}
\end{figure}
\par
\begin{figure}[t]
\begin{center}
\leavevmode
\epsfxsize=300pt
\epsfbox{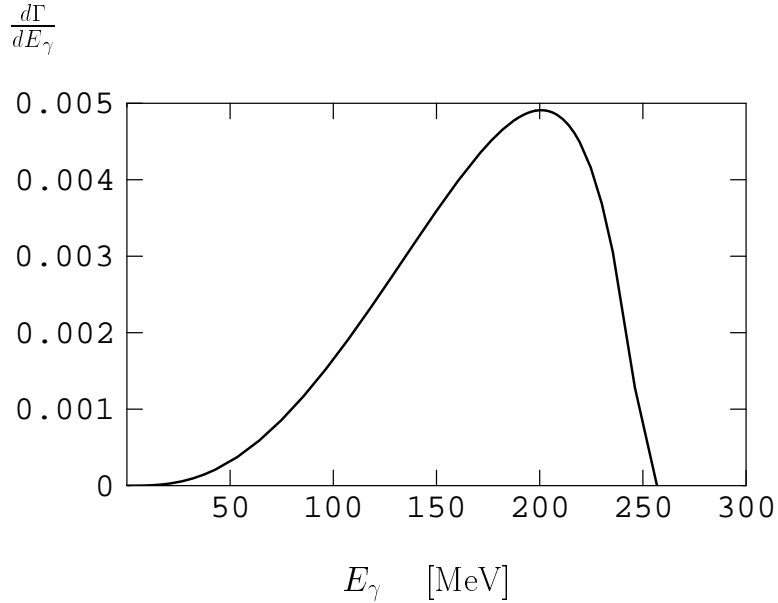}    
\caption{Spectrum of the photon energy $E_\gamma$.}
\label{fig:e2pggsp2}
\end{center}
\end{figure}

    Calculated spectrum of the photon invariant mass square 
$m_{\gamma\gamma}^{2}$ for the \epg decay is shown in Fig \ref{fig:e2pggsp1}. 
As this spectrum is compared with those calculated by ChPT in \cite{Ko}, 
we find ours to be similar to the one for 
$d_{3}=4.5 \times 10^{-2}$  GeV$^{-2}$ 
in \cite{Ko} which involves an additional $O(p^{6})$ contribution to 
the original lagrangian.
Spectrum of the photon energy $E_{\gamma}$ for the \epg decay is shown in
Fig \ref{fig:e2pggsp2}, and given in \cite{Ame} in ChPT.
Both are also similar, though there is no experimental result.
\par
    In our calculation of the \epg decay, we evaluate only the quark-box 
diagram given in Fig \ref{fig:e2pggf}. Since the vector and axialvector 
four-quark interactions are not included in our model, the only other 
contribution to this process is the scalar resonance exchange. In the ENJL 
model the contribution of the scalar resonance exchange is small\cite{Bel}. 
The relevant process is $\eta \to a_0 \pi^0 \to \gamma \gamma \pi^0$.
Since the $a_0 \to \gamma \gamma$ decay amplitude is known to be small 
experimentally, the scalar resonance contribution is suppressed.
Although the effect of the scalar channel quark-antiquark correlation is not 
taken into account in our calculation, the scalar resonance contribution is 
(partly) included in the box diagram.
\par
    If one includes the vector and axialvector four-quark interaction in the 
NJL model, the pseudoscalar meson properties are affected through the 
pseudoscalar-axialvector channel mixing and the model parameters with and
without the vector and axialvector four-quark interaction are different.
We expect that the models with and without the vector-axialvector interaction
predict similar results for the processes involving only the pseudoscalar
mesons with energies much below the vector meson masses. It is further
argued that the contribution of the quark-box diagram to the $\gamma \gamma
\to \pi^0 \pi^0$ process, that is similar to \epg, is quite close to that of
the vector meson exchange in the vector dominance model\cite{Bijnens2}.    
\section{Scalar Quark Contents in Nucleon}
\hspace*{\parindent}The $U_A(1)$ breaking six-quark flavor-determinant 
interaction ${\cal L}_6$ given in Eq. (\ref{njl4}) gives rise to flavor 
mixing not only in the pseudoscalar channel but also in the scalar channel. 
Therefore, it is important to study the effects of the $U_A(1)$ breaking 
interaction in the scalar $\bar qq$ channels. 
\par
    In the NJL model, it is known that the masses of the scalar mesons become 
just the twice of the constituent quark mass in the chiral limit.  
Introducing the explicit breaking of the chiral symmetry by the current 
quark masses pushes up the scalar meson masses above the quark-antiquark
threshold.  Since the NJL model does not confine quarks, we do not apply it 
to the scalar mesons in this article.
\par
    In this section we will discuss the scalar quark contents in nucleon as 
well as the pion-nucleon sigma term $\Sigma_{\pi N}$ and the kaon-nucleon 
sigma term $\Sigma_{KN}$.  We use a rather naive additive quark model for 
the nucleon state, namely, the nucleon is made up by three noninteracting 
constituent quarks whose masses are determined by the gap equation shown in 
Eq. (\ref{gap}).  The scalar quark content of flavor a in the proton is then 
obtained as follows.
\be
\langle P | \bar q^a q^a | P \rangle = 2 \langle U | \bar q^a q^a | U \rangle
+ \langle D | \bar q^a q^a | D \rangle \, ,
\ee 
with $| U \rangle$ ($| D \rangle$) is the constituent u- (d-) quark state.  
The amount of the $\bar qq$ content of flavor $a$ in a constituent quark Q of 
flavor $b$ is deduced by using the Feyman-Hellman theorem, i.e., 
\be
\langle Q^b | \bar q^a q^a | Q^b \rangle = \frac{\partial M_b}{\partial m_a}
\, ,
\ee
where $m_a$ and $M_b$ are the current and constituent quark masses, 
respectively.  Here the isospin symmetry is assumed and therefore the 
following relations hold.  
$\langle U | \bar uu | U \rangle = \langle D | \bar dd | D \rangle$, 
$\langle U | \bar dd | U \rangle = \langle D | \bar uu | D \rangle$ and
$\langle U | \bar ss | U \rangle = \langle D | \bar ss | D \rangle$.
\par
\begin{figure}[t]
\begin{center}
\leavevmode
\epsfxsize=300pt
\epsfbox{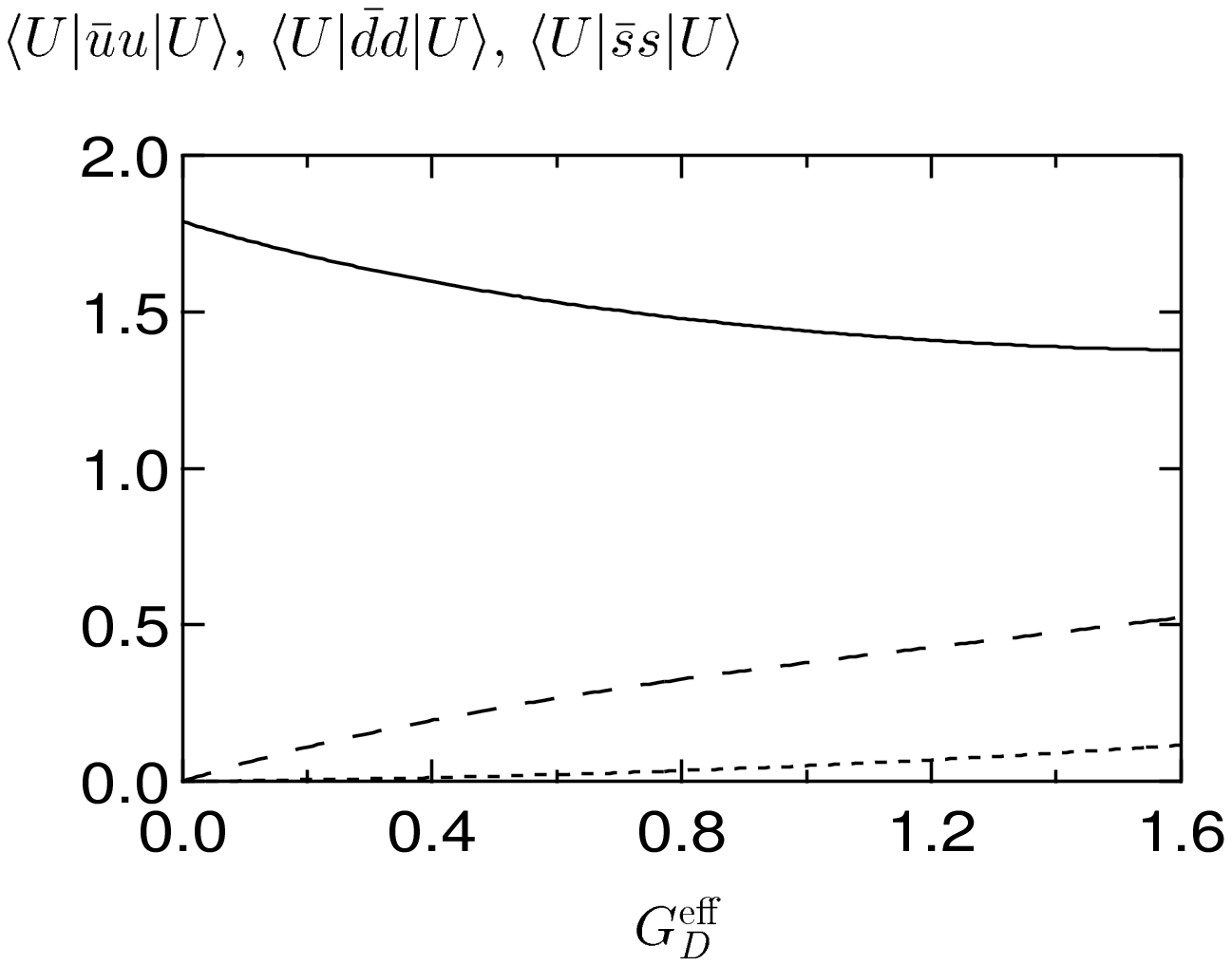}    
\caption{Dependence of the scalar quark contents in the constituent u-quark
on the dimension-less coupling constant $G_D^{\rm eff}$.  The solid line, 
dashed line and dotted line indicate $\langle U | \bar uu | U \rangle$, 
$\langle U | \bar dd | U \rangle$ and $\langle U | \bar ss | U \rangle$, 
respectively.}
\label{fig:qcont}
\end{center}
\end{figure}

    We have calculated the scalar quark contents in the 
constituent u-quark as functions of $G_D^{\rm eff}$ and the results are 
shown in Fig. \ref{fig:qcont}.
At $G_D^{\rm eff} = 0$, both $\langle U | \bar dd | U \rangle$ and 
$\langle U | \bar ss | U \rangle$ vanish as expected.  
Because of the larger mass of the strange quark, 
$\langle U | \bar ss | U \rangle$ is strongly suppressed compared to 
$\langle U | \bar dd | U \rangle$.  
At $G_D^{\rm eff} = 0.7$, our results are 
\be 
\langle P | \bar uu | P \rangle = 3.30 \,, \quad 
\langle P | \bar dd | P \rangle = 2.10 \,, \quad
\langle P | \bar ss | P \rangle = 0.08 \,,
\ee
and therefore, the strange quark content in the proton is rather small,
\be
y \equiv \frac{2\, \langle P |\, \bar ss \, | P \rangle}
{\langle P | \, \bar uu + \bar dd \, | P \rangle} = 0.03 \, .
\ee 
This value is smaller than the  ``standard'' value $y \approx 0.2$. 
However, as we will see later, our result depends on the choice of the 
current u,d-quark mass and therefore should not be taken seriously. 
\par
\begin{figure}[t]
\begin{center}
\leavevmode
\epsfxsize=300pt
\epsfbox{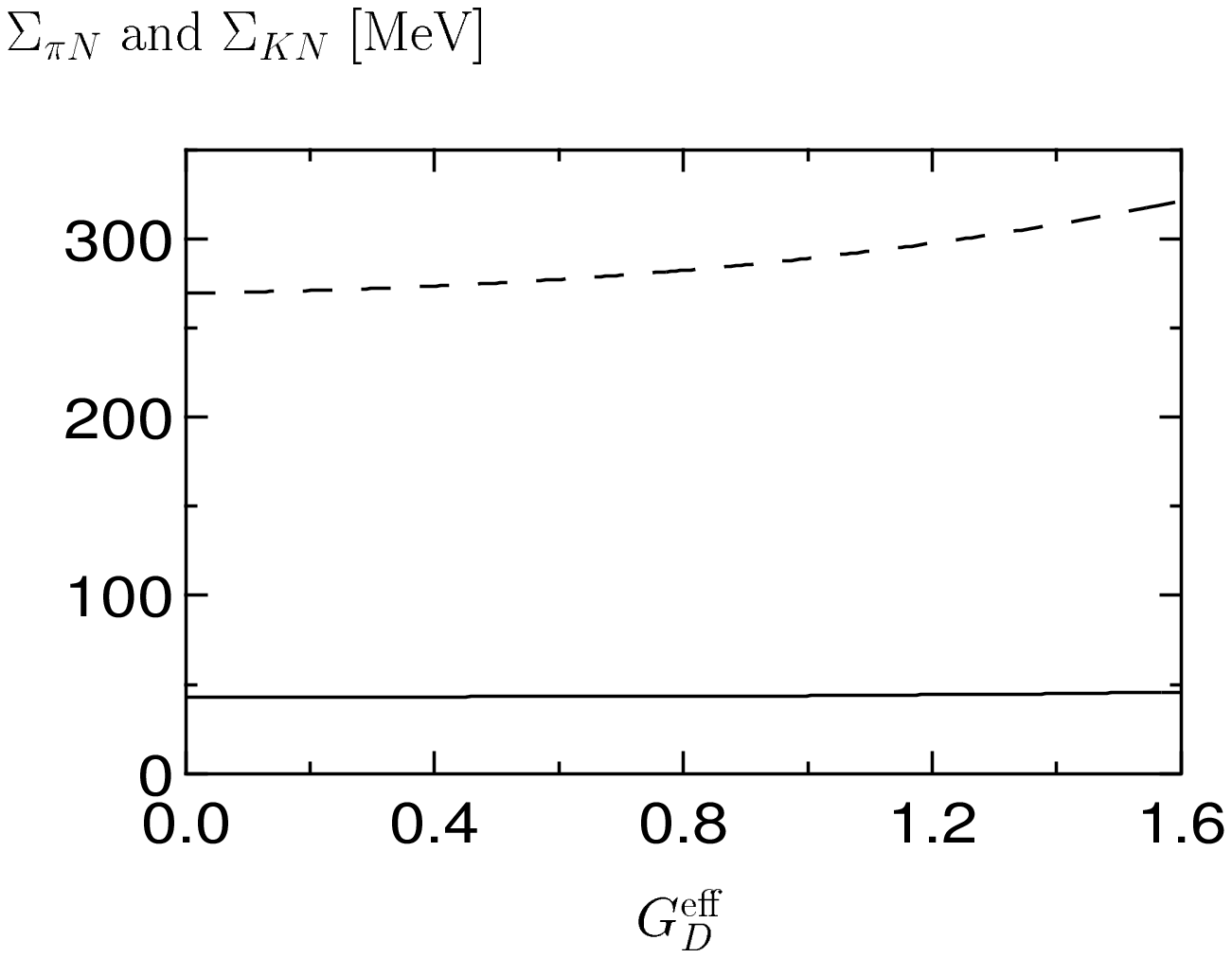}    
\caption{Dependence of the $\Sigma_{\pi N}$ and $\Sigma_{KN}$ terms
on the dimension-less coupling constant $G_D^{\rm eff}$.  The solid line 
indicates $\Sigma_{\pi N}$ term and the dashed line indicates 
$\Sigma_{K N}$ terms, respectively.}
\label{fig:sigma}
\end{center}
\end{figure}

    Assuming the isospin symmetry, the $\Sigma_{\pi N}$ and $\Sigma_{KN}$
terms are represented using the scalar quark contents in the nucleon as
follows.
\bea
\Sigma_{\pi N} & = & \hat m \,
\langle N |\, \bar uu + \bar dd \, | N \rangle \, , \\
\Sigma_{KN} & = & \frac{3}{2} (\hat m + m_s)\, 
\langle N |\, \frac{1}{2} ( \bar uu + \bar dd) + \bar ss \,| N \rangle \, , 
\eea
with $\hat m = (m_u + m_d)/2$. The calculated $\Sigma_{\pi N}$ and 
$\Sigma_{KN}$ terms are shown in Fig. \ref{fig:sigma}.
The $\Sigma_{\pi N}$ term is almost independent on $G_D^{\rm eff}$.  
At $G_D^{\rm eff} = 0.7$, we obtained $\Sigma_{\pi N} = 43.2$ MeV which is 
in good agreement with the value 
extracted from the low-energy $\pi$N scattering data \cite{GLS}:
$\Sigma_{\pi N} = (45 \pm 10)$ MeV.  As for the  $\Sigma_{K N}$ term, 
our result at $G_D^{\rm eff} = 0.7$ is 280 MeV.  
\par
    In the additive quark assumption of the nucleon states, the following 
relation holds.
\be
\langle P \, | \, 2 \bar dd - \bar uu \, | \, P \rangle = 3 
\langle U \, | \, \bar dd \, | \, U \rangle \, . 
\ee
Therefore, it vanishes if there exists no flavor mixing in the scalar 
channel.  Our result at $G_D^{\rm eff} = 0.7$ is 
$\langle P \, | \, 2 \bar dd - \bar uu \, | \, P \rangle = 0.89$. 
\par
    The pion-nucleon sigma term and the scalar quark contents in the 
nucleon are extensively studied in the three-flavor NJL model in 
Ref. \cite{BJM88}.  As pointed out there, the scalar quark contents in 
the constituent quarks depend on the current quark masses nonlinearly. 
In order to study this nonlinearity, we change the current u,d-quark mass 
from 7.5 MeV to 8.5 MeV.  Other four parameters of the model ($m_s$, $G_S$
$G_D$ and $\Lambda$) are determined so as to reproduced the observed values 
of $m_\pi$, $m_K$, $m_\eta$ and $f_\pi$. 
Thus the size of the chiral symmetry breaking is fixed though the current 
u,d-quark mass is changed. 
The calculated results are shown in Table \ref{tab:qcont}. 
\begin{table}
\caption{The scalar quark contents in the constituent u-quark and the 
$\Sigma_{\pi N}$ and $\Sigma_{KN}$ terms in the three different current 
u,d-quark masses.}
\label{tab:qcont}
\begin{center}
\small
\begin{tabular}{|c|ccc|cc|}
\hline
\col{$m_{u,d}$}\,[MeV] & \col{$\langle U \, |\, \bar uu \,|\, U \rangle$} & 
\col{$\langle U \, |\, \bar dd \, |\,  U \rangle$}& 
\col{$\langle U \, |\, \bar ss \, |\,  U \rangle$}&
\col{$\Sigma_{\pi N}$} \, [MeV] & \col{$\Sigma_{K N}$} \, [MeV] \\
\hline
7.5 & 1.582 & 0.707 & 0.210 & 51.5 & 390.6 \\
8.0 & 1.388 & 0.478 & 0.091 & 44.8 & 308.8 \\
8.5 & 1.232 & 0.300 & 0.009 & 39.1 & 243.0 \\
\hline
\end{tabular}
\end{center}
\end{table}
The scalar quark contents in the constituent u-quark and the sigma terms, 
especially, $\langle U \, | \, \bar ss \, | \, U \rangle$ and $\Sigma_{KN}$
depend on the current u,d-quark mass rather strongly. 
\par
    We next discuss the validity of the additive quark assumption for the 
nucleon state.  Kunihiro and Hatsuda \cite{KH90, HK94} studied the effects of 
the confinement and the short-range spin-spin interaction between the 
constituent quarks in baryons in the framework of the nonrelativistic 
potential model.  They have found that the effects of the residual 
interactions between the constituent quarks in the proton increase the 
$\bar uu$, $\bar dd$ and $\bar ss$ contents of the proton by about 
5\%, 24\% and 13\%, respectively.  The contributions of the kinetic term of
the confined quark are flavor independent and negative, while those of the 
short-range spin-spin interaction are flavor dependent and positive. 
The flavor-mixing effect by the short-range spin-spin interaction is rather 
large.  For simplicity, we consider the 
$\langle U \, | \, \bar dd \, | \, U \rangle = 
 \langle U \, | \, \bar ss \, | \, U \rangle = 0$ case.  Using the expression 
of $\langle P \, | \, \bar q_i q_i \, | \, P \rangle$ given in 
Ref. \cite{HK94}, one obtains 
\be
\langle P \, | \, 2 \bar dd - \bar uu \, | \, P \rangle = \frac{6 b}{M_u^3} \,
\langle U \, | \, \bar uu \, | \, U \rangle  \, ,
\ee
where $b = (176.4 {\rm MeV})^3$ is the strength of the short-range 
spin-spin interaction.  Inserting our numerical results at 
$G_D^{\rm eff} =0$, i.e., $M_u = 325$ MeV and 
$\langle U \, | \, \bar uu \, | \, U \rangle = 1.79$, we get 
$\langle P \, | \, 2 \bar dd - \bar uu \, | \, P \rangle = 1.72$. 
\par
    Recently, the static properties of the nucleon have been studied in 
the relativistic Faddeev approach using the two-flavor NJL model \cite{AIBY}.
Their results of the scalar quark contents in the nucleon are 
$\langle P \, | \, \bar uu \, | \, P \rangle = 1.795$ and 
$\langle P \, | \, \bar dd \, | \, P \rangle = 1.095$. 
Since the single quark renormalization factor $\partial M/\partial m$ is 
not included in their calculations, one should compare above numbers 
with those in the additive quark model, i.e., 
$\langle P \, | \, \bar uu \, | \, P \rangle = 2$ and 
$\langle P \, | \, \bar dd \, | \, P \rangle = 1$.  
The residual interactions between quarks in the nucleon decrease the u-quark
content by about 10\% and increase the d-quark content by about 10\%. 
It is a contrast to the results in the nonrelativistic quark model.  
As for the flavor mixing effect, one gets 
$\langle P \, | \, 2 \bar dd - \bar uu \, | \, P \rangle = 0.395 
\langle U \, | \, \bar uu \, | \, U \rangle \simeq 0.71$.  Here we have used 
our result: $\langle U \, | \, \bar uu \, | \, U \rangle = 1.79$.  
In the nonrelativistic quark model approach, the flavor mixing effect 
entirely comes from the short-range spin-spin interaction.  On the other 
hand, in the relativistic Faddeev approach, only the scalar diquark state 
is included.  Inclusion of the axialvector diquark may be important 
for the flavor mixing in the scalar quark contents in the nucleon.   
\par
    In the above two approaches, the effects of the pion cloud around the 
quark-core of the nucleon is not taken into account.  Wakamatsu has 
studied the scalar quark contents in the nucleon using the chiral quark
soliton model as functions of the constituent u,d-quark mass \cite{W92}.
His results are as follows.  At $M_{u,d}=350$MeV, 
$\langle P \, | \, \bar uu \, | \, P \rangle = 1.807$, 
$\langle P \, | \, \bar dd \, | \, P \rangle = 1.223$ and 
at $M_{u,d}=450$MeV, 
$\langle P \, | \, \bar uu \, | \, P \rangle = 1.382$, 
$\langle P \, | \, \bar dd \, | \, P \rangle = 0.978$. 
As discussed in \cite{W92}, 
the flavor-asymmetry of the sea-quark in the nucleon gives rise to the 
flavor mixing phenomena and the calculated results are 
$\langle P \, | \, 2 \bar dd - \bar uu \, | \, P \rangle = 0.639$ at 
$M_{u,d}=350$MeV and 
$\langle P \, | \, 2 \bar dd - \bar uu \, | \, P \rangle = 0.574$ at 
$M_{u,d}=450$MeV.  Since the chiral quark soliton model do not have 
the dynamics of the spontaneous breaking of chiral symmetry, the effects of 
the single quark renormalization factor $\partial M/\partial m$ is not 
taken into account.  
\par
    In conclusion, as is expected, the flavor mixings in the scalar channel 
are approximately proportional to the strength of the $U_A(1)$ breaking 
interaction in the NJL model.  However their magnitudes depend on the 
choice of the model parameters.  Furthermore, the scalar quark contents 
in the proton have many origins, and therefore we cannot draw a definite 
conclusion on the effects of the $U_A(1)$ breaking interaction. 
\section{Effects of the $U_A(1)$ Anomaly in Baryons}
\hspace*{\parindent}Since the effects of the $U_A(1)$ anomaly are rather 
large in the pseudoscalar sector, it is natural to ask if one can see some 
effects in the baryon sector.  It was pointed out in \cite{SR89} that the 
instanton can play an important role in the description of spin-spin forces, 
particularly for light baryons.  The pattern of these effects can be very hard 
to disentangle from one-gluon exchange. The effects of the instanton induced 
interaction in baryon number $B=2$ systems were studied in \cite{OT89}.
It was shown that an attraction between two nucleons is obtained by
the two-body 
instanton induced interaction, while the three-body interaction is strongly 
repulsive in the H-dibaryon channel and makes the H-dibaryon almost unbound.
\par
   We estimate the effects of the $U_A(1)$ anomaly on the $B = 1$ and 
$B = 2$ systems by employing the six-quark determinant interaction given in 
Eq. (\ref{njl4}) whose strength was determined so as to reproduce the 
observed $\eta$-meson mass, the \egg decay width and the \epg decay width, 
namely, $G_D^{\rm eff} = 0.7$.  It is done by calculating the matrix 
elements of the the $U_A(1)$ breaking interaction hamiltonian with respect to 
unperturbed states of the MIT bag model and the nonrelativistic quark model 
(NRQM).  For $B = 2$ systems, we only consider the $(0S)^6$ configuration 
of the six valence quark states. Therefore, the matrix element with respect 
to such a state gives a measure of the contribution of the $U_A(1)$ breaking 
interaction either to the dibaryon or to the short-range part of the 
interaction between two baryons.  The determinant interaction induces not 
only three-body but also two-body interactions of valence quarks when the 
vacuum has a nonvanishing quark condensate.  The details of the 
calculation are described in \cite{MT93}.  
\begin{table}
\caption{Contribution of the two-body term to octet and decuplet baryons. 
All the entries are in units of MeV.}
\label{tab:2btb1}
\begin{center}
\small
\begin{tabular}{|c|cccc|cccc|}
\hline
wave function & \col{$N$} & $\Sigma$ & $\Xi$ & $\Lambda$ 
& $\Delta$ & ${\Sigma ^*}$ & ${\Xi^*}$ & $\Omega$ \\
\hline
\col{MIT} & $-43.9$ & $-41.2.$ & $-41.2$ & $-42.9$ & $0$ &
$0.12$ & $0.12$ & $0$ \\
\col{NRQM} & $-40.88$ & $-36.6$ & $-36.6$ & $-39.4$ & $0$ &
$0.07$ & $0.07$ & $0$ \\
\hline
\end{tabular}
\end{center}
\end{table}
\begin{table}
\caption{Baryon component, $SU(3)$ multiplet, spin, isospin and strangeness 
of the eight channels of two octet baryons.}
\label{tab:channel}
\begin{center}
\small
\begin{tabular}{|c|ccccc|}
\hline
channel & \col{Baryon component} & $SU(3)$ multiplet & Spin & Isospin & 
Strangeness \\
\hline
\col{I} & $NN$ & {\bf 10*} & 1 & 0 & 0 \\
\col{II} & $NN$ & {\bf 27} & 0 & 1 & 0 \\
\hline
\col{III} & $N\Sigma$ & {\bf 27} & 0 & 3/2 & -1 \\
\col{IV} & $N\Sigma-N\Lambda$ & {\bf 27} & 0 & 1/2 & -1 \\
\col{V} & $N\Sigma-N\Lambda$ & {\bf 10*} & 1 & 1/2 & -1 \\
\col{VI} & $N\Sigma$ & {\bf 10} & 1 & 3/2 & -1 \\
\col{VII} & $N\Sigma-N\Lambda$ & {\bf 8} & 1 & 1/2 & -1 \\
\hline
\col{VIII} & $H$ & {\bf 1} & 0 & 0 & -2 \\
\hline
\end{tabular}
\end{center}
\end{table}

\par
    Table \ref{tab:2btb1} shows the contribution of 
the two-body term for $B = 1$. 
The contribution to the decuplet baryons vanishes in the $SU(3)$ limit 
and therefore comes only from the $SU(3)$ asymmetry of the quark wave
function. 
The three-body term does not contribute to the $B = 1$ states. 
Thus the  N$\Delta$ mass difference due to the $U_A(1)$ 
breaking interaction is about 15\% of the observed one.   
\par
    We next discuss the case of $B = 2$.  We consider all the possible 
channels which are made of two octet baryons listed in Table \ref{tab:channel}.
Table \ref{tab:2btb2} shows the contribution of the two-body term.  
The channel VIII gets 
the strongest attraction, about $170$ MeV, and the channel VII gets the second 
strongest attraction.  The contributions of the three-body term to the 
H-dibaryon and strangeness $-1$ channels are given in Table \ref{tab:3btb2}.
It should be noted that the three-body term has no effect on the NN
channels, and that the contributions to the channels III, IV and V
reflect the $SU(3)$ breaking in the quark wave function.
The contributions of the three-body term in channels VI, VII and VIII are 
remarkable and one will be able to observe some effects experimentally.  
\begin{table}
\caption{Contributions of the two-body term to the eight channels of two 
octet baryons listed in Table 3.  All the entries are in units of MeV.}
\label{tab:2btb2}
\begin{center}
\small
\begin{tabular}{|c|cc|ccccc|c|}
\hline
wave function & \col{I} &  II & III & IV & V & VI & VII & VIII \\
\hline
\col{MIT} & $-90$ & $-85$ & $-89$ & $-86$ & $-94$ & $-97$
& $-122$ & $-163$ \\
\col{NRQM} & $-120$ & $-105$ & $-102$ & $-104$ & $-118$ & $-117$
& $-148$ & $-183$ \\
\hline
\end{tabular}
\end{center}
\end{table}
\begin{table}
\caption{Contributions of the three-body term to the H-dibaryon (VIII) and 
strangeness $-1$ two octet baryon channels (III-VII).  All the entries are 
in units of MeV.}
\label{tab:3btb2}
\begin{center}
\small
\begin{tabular}{|c|ccccc|c|}
\hline
wave function & \col{III} & IV & V & VI & VII & VIII \\
\hline
\col{MIT} & $-6\times 10^{-2}$ & $-6\times 10^{-2}$ &
$-7\times 10^{-2}$ & $20.7$ & $25.1$ & $40.7$ \\
\col{NRQM} & $-5\times 10^{-2}$ & $-5\times 10^{-2}$ &
$-5\times 10^{-2}$ & $28.3$ & $34.9$ & $56.1$ \\
\hline
\end{tabular}
\end{center}
\end{table}
\par
    We should comment on the difference between the determinant 
interaction  used here and the instanton-induced interaction 
used in ref. \cite{OT89}.  The relative contributions of the $U_A(1)$ 
breaking interaction within the baryonic sector or within the mesonic 
sector are similar for the two interactions. However, the ratio of
those in the baryonic sector to those in the mesonic sector is about
$\frac{4}{7}$.  
Namely, if one fixes the strength of the interaction so as to give the 
same mass difference of $\eta$ and $\eta'$, the effects of the
instanton-induced interaction in the baryonic sector would be about
$\frac{7}{4}$ stronger than those of the determinant interaction.  
After this correction the strength of the present $U_A(1)$ breaking
interaction is consistent with that used in the calculation of the
baryon-baryon interaction in ref. \cite{OT89}.
\section{Summary and conclusions}
\hspace*{\parindent}Using an extended three-flavor NJL model that includes 
the 't Hooft instanton induced $U_A(1)$ breaking interaction, we have 
studied the \egg, \egm, \ege, and \epg decays as well as the $\eta$-meson 
mass, $\eta$ decay constant and flavor $SU(3)$ singlet-octet mixing angle 
$\theta$ of the $\eta$ meson.  Advantages of our approach are 
as follows.  (1) Effects of the explicit chiral symmetry breaking by the 
current quark masses and the $U_A(1)$ anomaly can be evaluated consistently 
on the $\eta$-meson decay amplitudes.
(2) One can study how the $\eta$-meson properties change when the strength 
of the $U_A(1)$ breaking interaction is changed.
\par
    We have found that the $\eta$-meson mass, the \egg, \egm and \epg decay 
widths are in good agreement with the experimental values when the $U_A(1)$
breaking is strong and the mixing angle $\theta$ is about zero.  
\par
    The calculated $\eta$ decay constant is almost independent of the 
strength of the $U_A(1)$ breaking interaction and is close to the pion decay 
constant.  It indicates that the $\eta$ meson does not lose the the 
Nambu-Goldstone boson nature though its mass and mixing angle are strongly 
affected by the $U_A(1)$ breaking interaction.
\par
    Our result of the mixing angle $\theta$ is about zero which is different 
from the value $\theta \simeq -20^{\circ}$ obtained in the PCAC + ABJ 
anomaly approach and ChPT.  We have discussed the possible origin of this 
difference in Sect. 4.2.  It should be stressed here that the 
$\eta' \to \gamma \gamma$ decay width is used to obtain the mixing angle in 
the PCAC and ChPT approaches.  However, since the $\eta'$ meson is heavy, 
it is rather questionable to study the $\eta' \to \gamma \gamma$ 
decay in the PCAC and ChPT approaches. 
\par
   Since the 't Hooft instanton induced $U_A(1)$ breaking interaction 
gives rise to the flavor mixing not only in the pseudoscalar $\bar qq$ 
channels but also in the scalar $\bar qq$ channels, we have studied the 
scalar quark contents in the nucleon, the pion-nucleon and the kaon-nucleon 
sigma terms.  The calculated pion-nucleon sigma term is almost independent 
of the strength of the $U_A(1)$ breaking interaction and in good agreement 
with the value extracted from the low-energy $\pi N$ scattering data.  
Concerning the flavor mixing effects, we have found that the amount of the
flavor mixing in the scalar quark contents in the nucleon depends on the 
strength of the spontaneous breaking of chiral symmetry and the residual 
interactions between the constituent quarks in the nucleon rather strongly.
In this sense, the scalar quark contents of the nucleon are interesting and 
important quantities.  Further studies are necessary. 
\par
    We have estimated the effects of the $U_A(1)$ anomaly on the baryon 
number $B=1$ and $B=2$ systems too.  We have found that the $N \Delta$ 
mass difference due to the $U_A(1)$ breaking interaction is about 15\%
of the observed one and the three-body term of the $U_A(1)$ breaking 
interaction gives 40-50 MeV repulsion in the H-dibaryon channel.
\par
    Finally, we should note that the NJL model does not confine quarks.  
Since the Nambu-Goldstone bosons, $\pi$, $K$ and $\eta$ are strongly bound, 
the NJL can describe their properties fairly well.  However the $\eta'$-meson
state in the NJL model has an unphysical decay of $\eta' \to q \bar q$. 
Therefore, we do not apply our model to the $\eta'$ meson. 
In order to study the role of the $U_A(1)$ anomaly on the low-energy QCD
further, the studies of the $\eta'$-meson properties are desirable.  
\newpage
\end{document}